\documentclass[AMA,Times1COL]{WileyNJDv5}

\articletype{}

\makeatletter
\renewcommand{\oddhead@titlepage@info}{}
\renewcommand{\evenhead@titlepage@info}{}
\renewcommand{\oddfoot@titlepage@info}{}
\renewcommand{\evenfoot@titlepage@info}{}
\makeatother

\usepackage{hyperref}
\hypersetup{
    colorlinks,
    linkcolor={red!50!black},
    citecolor={blue!50!black},
    urlcolor={blue!80!black}
}

\usepackage{lineno}

\begin{document}

\title{A primer on inference and prediction with epidemic renewal models and sequential Monte Carlo}

\author[1]{Nicholas Steyn}
\author[2]{Kris V. Parag}
\author[3]{Robin N. Thompson}
\author[1,4]{Christl A. Donnelly}

\address[1]{\orgdiv{Department of Statistics}, \orgname{University of Oxford}, \orgaddress{\state{Oxford}, \country{UK}}}

\address[2]{\orgdiv{MRC Centre for Global Infectious Disease Analysis}, \orgname{Imperial College London}, \orgaddress{\state{London}, \country{UK}}}

\address[3]{\orgdiv{Mathematical Institute}, \orgname{University of Oxford}, \orgaddress{\state{Oxford}, \country{UK}}}

\address[4]{\orgdiv{Pandemic Sciences Institute}, \orgname{University of Oxford}, \orgaddress{\state{Oxford}, \country{UK}}}

\corres{Corresponding author Nicholas Steyn. \email{nicholas.steyn@univ.ox.ac.uk}}

\abstract[Abstract]{Renewal models are widely used in statistical epidemiology as semi-mechanistic models of disease transmission. While primarily used for estimating the instantaneous reproduction number, they can also be used for generating projections, estimating elimination probabilities, modelling the effect of interventions, and more. We demonstrate how simple sequential Monte Carlo methods (also known as particle filters) can be used to perform inference on these models. Our goal is to acquaint a reader who has a working knowledge of statistical inference with these methods and models and to provide a practical guide to their implementation. We focus on these methods’ flexibility and their ability to handle multiple statistical and other biases simultaneously. We leverage this flexibility to unify existing methods for estimating the instantaneous reproduction number and generating projections. A companion website \textit{\href{https://nicsteyn2.github.io/SMCforRt/}{SMC and epidemic renewal models}} provides additional worked examples, self-contained code to reproduce the examples presented here, and additional materials.}



\maketitle


\section{Introduction}

Modern epidemiology relies on statistical models to track and project the spread of infectious diseases, inform public health policymaking, and estimate disease burden \cite{lewnardEmergingChallengesOpportunities2019}. A commonly used model for these purposes is the semi-mechanistic renewal model, which relates past incidence to current incidence through a simple renewal process \cite{diekmannLimitingBehaviourEpidemic1977, fraserEstimatingIndividualHousehold2007, bhattSemimechanisticBayesianModelling2023}. While this model is most commonly used for estimating the instantaneous reproduction number $R_t$ \cite{coriNewFrameworkSoftware2013, abbottEstimatingTimevaryingReproduction2020,paragImprovedEstimationTimevarying2021}, it can also be used for generating projections \cite{banholzerComparisonShorttermProbabilistic2023, bracherPreregisteredShorttermForecasting2021,abbottEstimatingTimevaryingReproduction2020, nouvelletSimpleApproachMeasure2018}, estimating elimination probabilities \cite{paragDecipheringEarlywarningSignals2021, djaafaraQuantitativeFrameworkDefining2021, thompsonUsingRealtimeModelling2024}, and modelling the effect of non-pharmaceutical interventions \cite{flaxmanEstimatingEffectsNonpharmaceutical2020, haugRankingEffectivenessWorldwide2020}, for example. Renewal models are frequently modified to account for various biases and limitations in epidemiological data \cite{nashRealtimeEstimationEpidemic2022, azmonEstimationReproductionNumber2014, paragAngularReproductionNumbers2023, paragQuantifyingInformationNoisy2022}. These modifications often necessitate bespoke methods for fitting the model to data, which can be difficult to implement and become increasingly complex with additional modifications.

Reviews of reproduction number estimation highlight key practical considerations to bear in mind when fitting the renewal model to epidemiological data. Gostic et al. (2020) \cite{gosticPracticalConsiderationsMeasuring2020} highlight generation interval misspecification, reporting delays, right-truncation due to observation delays, incomplete observation, and smoothing windows as particular concerns. Nash, Nouvellet, \& Cori (2022) \cite{nashRealtimeEstimationEpidemic2022} identify 54 papers that make modifications to EpiEstim (arguably the most popular $R_t$ estimator) to account for such biases. They note delayed and missing reported case data, weekly reporting noise, alternative smoothing methods, imported cases, and temporal aggregation of reported data as common considerations. These adjustments are just as necessary for generating projections and other epidemiological analyses as they are for $R_t$ estimation.

In this primer, we present a pedagogical overview of how sequential Monte Carlo (SMC) methods \cite{doucetSequentialMonteCarlo2001, sarkkaBayesianFilteringSmoothing2013}, which are well established in other contexts, can be used as flexible and general-purpose tools for fitting renewal models to epidemiological data. The overall approach can be viewed as a distillation of the methods employed by Watson et al. (2024) \cite{watsonJointlyEstimatingEpidemiological2024}.  We focus on how these methods can handle multiple biases simultaneously and provide a unified framework for a range of tasks including reproduction number estimation and generating projections.

SMC methods, often called particle filters, are a class of algorithms that are used to fit general hidden-state models to observed data. They use a collection of samples (called particles) to represent the posterior distribution of ``hidden states'' at each time step (e.g. the instantaneous reproduction number or daily infection incidence) given some observed data (e.g. reported cases). These particles are projected forward according to a state-space transition model (typically an epidemic model of some kind) and weighted by an observation model (typically representing a noisy observation process). Such methods are particularly powerful, as they only require simulations from the epidemic model, rather than evaluations of a likelihood, allowing for simulation-based inference \cite{mckinleyInferenceEpidemicModels2009}. Accounting for the aforementioned delays and biases simply requires incorporating these in the state-space transition and observation models, allowing a wide range of models to be fit using a single framework.

Despite being well suited to fitting epidemiological models and having a relatively straightforward and intuitive implementation, and calls for their further use \cite{daiInvitationSequentialMonte2022}, SMC methods are not widely used when fitting epidemic renewal models. They have seen greater use in fitting compartmental models \cite{safarishahrbijariPredictiveAccuracyParticle2017, liRealTimeEpidemiologyAcute2024, weldingRealTimeAnalysis2019, weldingRealTimeAnalysis2019, storvikSequentialMonteCarlo2023, sheinsonComparisonPerformanceParticle2014, wonEstimatingInstantaneousReproduction2023, yangComparisonFilteringMethods2014}, while examples of their application to agent-based models \cite{rimellaApproximatingOptimalSMC2023}, Hawkes processes \cite{koyamaEstimatingTimevaryingReproduction2021}, and renewal models \cite{yangBayesianDataAssimilation2022, watsonJointlyEstimatingEpidemiological2024, plankEstimationEndofoutbreakProbabilities2025} are more limited. This discrepancy likely arises from the Markovian nature of many compartmental models, a property which is leveraged by many popular SMC methods, whereas the renewal model is explicitly non-Markovian. We are aware of two reviews of SMC methods in epidemiology \cite{temfackReviewSequentialMonte2024b,endoIntroductionParticleMarkovchain2019}, both of which are high quality and provide complementary information to this primer, though neither covers methods applicable to the renewal model. To aid comparison with these papers, a discussion of similarities and differences with this primer is provided \href{https://nicsteyn2.github.io/SMCforRt/smc-other.html}{online}.

Methods for fitting renewal models that feature comparable flexibility typically rely on Hamiltonian Monte Carlo (HMC)\cite{abbottEstimatingTimevaryingReproduction2020, scottEpidemiaModelingEpidemics, lisonGenerativeBayesianModeling2024} (as implemented in Stan \cite{standevelopmentteamStanReferenceManual2024}), often coupled with an approximate Gaussian process state-space model (such as in EpiNow2 \cite{abbottEstimatingTimevaryingReproduction2020}, which is available as a plug-and-play software package). However, HMC-based methods do not support discrete state spaces or parameters, can be comparatively computationally intensive, and cannot be updated online as new data are observed. Additionally, Gaussian process approximations introduce significant mathematical complexity. Our goal is to enable a researcher to rapidly implement their own methods, ensuring they understand the process from start to finish. An alternative flexible approach is the Laplacian P-spline method of EpiLPS \cite{gressaniEpiLPSFastFlexible2022}, though its fully Bayesian implementation is also gradient-based, again limiting it to continuous state spaces and parameters, and the mathematical complexity of this method is also high. Finally, approximate Bayesian computation methods could also be used to fit renewal models, although (as for SMC) the literature is largely focused on compartmental models \cite{kypraiosTutorialIntroductionBayesian2017, mckinleyApproximateBayesianComputation2018}.

We aim to familiarise the reader with both the discrete-time renewal model and SMC methods. We provide simple implementations of SMC-type algorithms, including a bootstrap filter with fixed-lag resampling and a particle marginal Metropolis-Hastings (PMMH) algorithm, for fitting renewal models to data. Their use is demonstrated in several epidemiologically interesting scenarios using national data from the COVID-19 pandemic in Aotearoa New Zealand. We also provide guidance on model evaluation and comparison. Our aim is not to provide a model for every scenario, but to equip the reader with the tools to construct and fit their own models.

This paper contains the technical details needed to construct and fit discrete-time epidemic renewal models with SMC methods, consolidating established methods into a single, accessible framework. Some readers may find it helpful to consult the example code online while reading, which is available as downloadable notebooks from \textit{\href{https://nicsteyn2.github.io/SMCforRt/}{SMC and epidemic renewal models}}. This website also contains additional examples, tutorials, and longer explanations of the topics considered here.

\section{Hidden-state models}

A typical hidden-state model consists of time-indexed unobserved hidden states $X_t$, observed data $y_t$, and parameter(s) $\theta$. These may be scalar or vector-valued. The hidden states represent the underlying process of interest (e.g. the true infection incidence and reproduction number) from which the observed data (e.g. reported cases) are generated. The goal is to infer the value of the hidden states $X_t$ and/or parameters $\theta$ given the observed data. SMC methods, discussed later, are a class of algorithms for fitting general hidden-state models to observed data \cite{doucetSequentialMonteCarlo2001}.

A hidden-state model is defined by two probability distributions. The first is the \textbf{state-space transition distribution}, which governs how hidden states evolve over time:
\[ P(X_t | X_{1:t-1}, \theta) \ \ \ \text{(state-space transition distribution)}\]
The notation $X_{s:t}$ denotes the values of this quantity at all time steps from $s$ to $t$, inclusive. In epidemiological applications, this distribution is typically implied by an epidemic model describing how the epidemic evolves over time. Stochastic compartmental models are a popular choice here \cite{temfackReviewSequentialMonte2024b, endoIntroductionParticleMarkovchain2019}, although other models (such as the renewal model) can be used.

The second probability distribution is the \textbf{observation distribution}, which relates observed data to the hidden states:
\[ P(y_t | X_{1:t}, y_{1:t-1}, \theta) \ \ \ \text{(observation distribution)}\]
This distribution can model the effect of various biases in the data (e.g. we may model underreporting using a binomial or beta-binomial distribution). Alternatively, non-mechanistic observation models (e.g. a Gaussian observation model) can be used to allow for general noise in the data.

\begin{figure}[h!]
    \centering
    \includegraphics[width=0.8\textwidth]{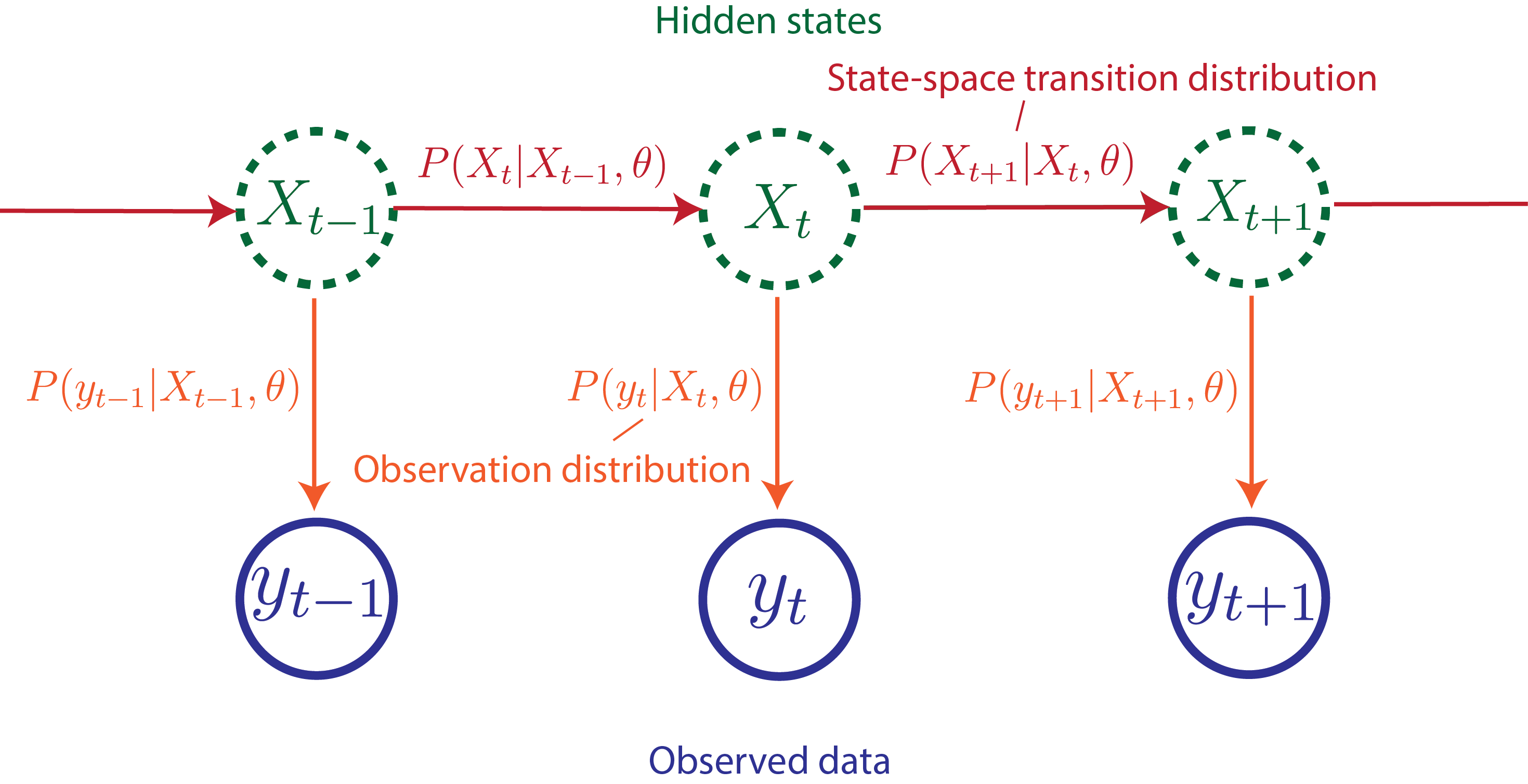}
    \caption{A diagram of a simple hidden-state model. The hidden states $X_t$ evolve over time according to the state-space transition distribution, from which observed data $y_t$ are generated via the observation distribution. Both distributions may depend on the parameter vector $\theta$.  In this example, hidden states $X_t$ depend only on the previous hidden state $X_{t-1}$, and observed data $y_t$ depend only on the hidden state at time $t$, making this particular model Markovian - an example of an HMM or POMP. In contrast, the more general non-Markovian models we focus on in this primer would require many more arrows in the diagram.}
    \label{fig:background-hiddenstatemodel}
\end{figure}

As these distributions model the state-space transition and observation processes, we also refer to \textit{state-space transition models} and \textit{observation models} throughout this primer. Together, they define a generative model of the epidemic, a simplified version of which is visualised in Figure \ref{fig:background-hiddenstatemodel}. This structure is particularly convenient in epidemiology: the underlying epidemic process forms the hidden states, which are assumed to drive the observed data.

Other names for hidden-state models include Hidden Markov Models (HMMs), when the model is Markovian \cite{kantasParticleMethodsParameter2015}; partially observed Markov processes (POMPs), also for Markovian models \cite{kingStatisticalInferencePartially2016}; and latent variable models, where $X_t$ are referred to as latent variables \cite{robertBayesianChoiceDecisionTheoretic2007}. The renewal model we examine is not Markovian, as infection incidence today depends on a convolution of infection incidence across multiple previous time steps, thus it cannot be treated as an HMM or POMP. Augmenting past states into $X_t$, i.e., setting $X_t = (\tilde{X}_t, \tilde{X}_{t-1}, \ldots)$ can transform a non-Markovian model into a Markovian one, at the cost of an enlarged hidden-state space. We do not consider this further in this primer.

Many popular epidemic models can be framed as a hidden-state model. EpiFilter, which can be used to estimate the instantaneous reproduction number $R_t$ \cite{paragImprovedEstimationTimevarying2021} or infer the elimination probability of an epidemic \cite{paragDecipheringEarlywarningSignals2021}, is explicitly formulated this way. In this model, $R_t$ is assumed to follow a Gaussian random walk (the state-space transition model) and observed cases $C_t$ are assumed to follow the Poisson renewal model (the observation model). A less obvious example of a hidden-state model is that by Fraser (2007) \cite{fraserInfluenzaTransmissionHouseholds2011}, which shares EpiFilter's observation model but assumes $R_t$ is piecewise constant on intervals of 10 days. Further examples include the various implementations of EpiNow2 \cite{abbottEstimatingTimevaryingReproduction2020}, and a model for temporally aggregated and underreported data by Ogi-Gittins et al. (2025)\cite{ogi-gittinsEfficientSimulationbasedInference2024}. These models all rely on the renewal model and employ bespoke inference methods, such as a grid-based approximation to the Bayesian filtering and smoothing equations in EpiFilter, maximum likelihood estimation in the model by Fraser et al. (2011) \cite{fraserInfluenzaTransmissionHouseholds2011}, or HMC in EpiNow2 \cite{abbottEstimatingTimevaryingReproduction2020}.

\section{The renewal model}

While SMC methods can be used to fit various hidden-state models, and many epidemic models can be formulated as hidden-state models, we focus on the discrete-time epidemic renewal model in this primer. The renewal model was introduced by Leonhard Euler in 1767 and popularised in the contemporary continuous-time formulation by Alfred Lotka in 1907, who used it to model age-structured population dynamics \cite{champredonEquivalenceErlangDistributedSEIR2018}. This continuous-time formulation was first introduced into epidemiology alongside the susceptible-infected-recovered (SIR) model by Kermack and McKendrick\cite{kermackContributionMathematicalTheory1927} and gained wider popularity during the 1970s \cite{diekmannLimitingBehaviourEpidemic1977}. The modern discrete-time formulation was proposed by Fraser (2007)\cite{fraserEstimatingIndividualHousehold2007} and subsequently popularised as a Bayesian estimator of $R_t$ in Cori et al. (2013) \cite{cauchemezEstimatingRealTime2006, coriNewFrameworkSoftware2013}. Since then, the discrete-time renewal model has become a popular semi-mechanistic model of disease transmission \cite{bhattSemimechanisticBayesianModelling2023}.

Denoting reported cases at time $t$ by $C_t$, the reproduction number by $R_t$, and the probability that a secondary case is reported $u$ time steps after the primary case by $\omega_u$ (often approximated by the probability mass function (PMF) of the serial interval), the renewal model relates reported cases at time $t$ to previously reported cases through the renewal equation:

\begin{equation}
    E[C_t] = R_t \sum_{u=1}^{u_{max}} C_{t-u} \omega_u \label{eq:renewalmodel}
\end{equation}

\begin{figure}[h!]
    \centering
    \includegraphics[width=\textwidth]{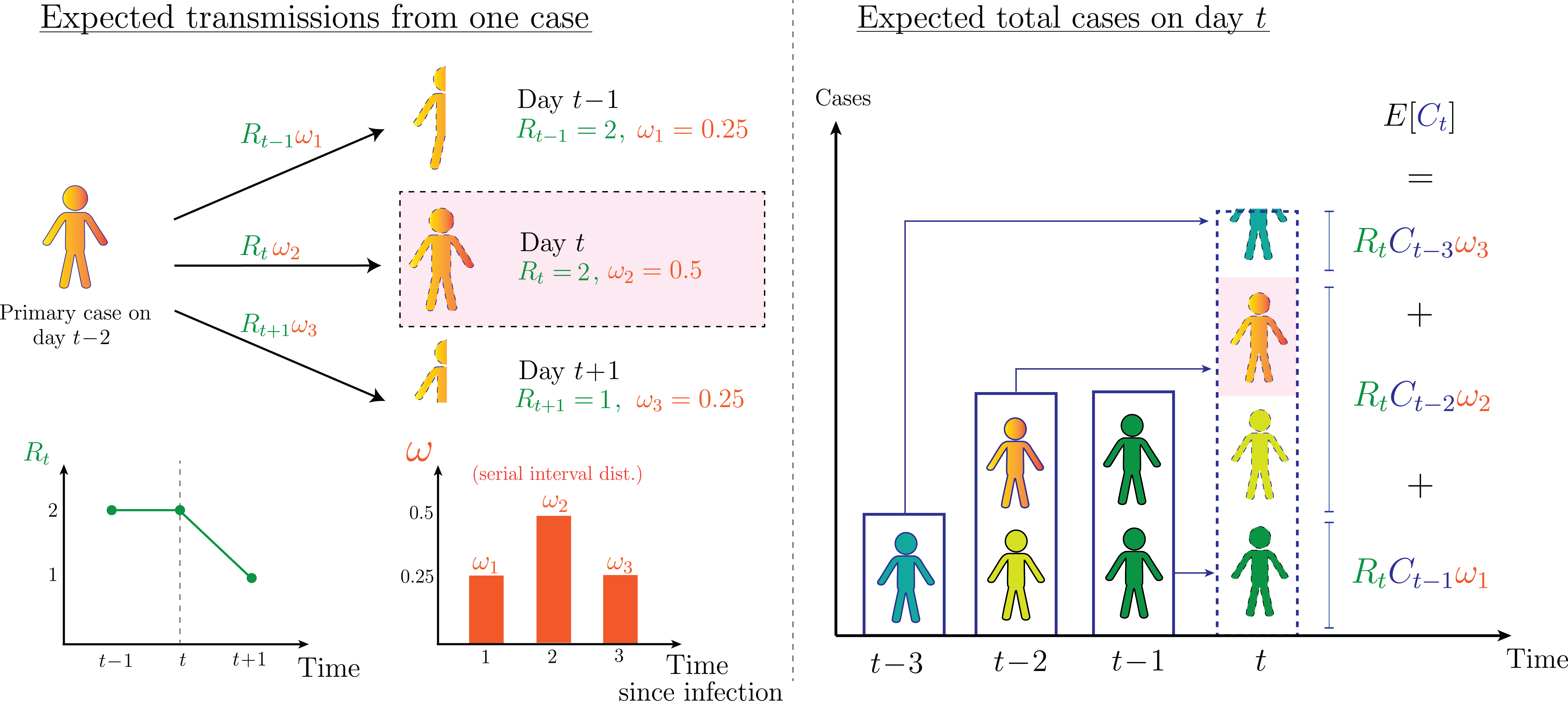}
    \caption{Diagram of the renewal model. In this example, the serial interval takes values $\omega_u = 0.25, 0.5, 0.25$ for $u = 1, 2, 3$ (a maximum serial interval of three time steps) and the reproduction number is assumed to take values $R_{t-1} = 2$, $R_{t} = 2$, and $R_{t+1} = 1$. On the left, the expected number of secondary cases produced by a primary case who was reported at time $t\!-\!2$ is shown (0.5 cases at time $t\!-\!1$, a single case at time $t$, and 0.25 cases at time $t\!+\!1$), with their expected contribution to total cases at time $t$ highlighted. On the right, the expected total cases at time $t$ is shown as the sum of the expected cases produced by primary cases reported $u = 1, 2, \text{ and } 3$ time steps ago, defining the renewal equation.}
    \label{fig:background-renewalmodel}
\end{figure}

If $C_t$ follows a Poisson distribution with mean $E[C_t]$, then we have the popular Poisson renewal model. However, other distributions can be used \cite{coriInferenceEpidemicDynamics2024}, often to account for overdispersion in transmission or ``superspreading'' \cite{lloyd-smithSuperspreadingEffectIndividual2005}. The serial interval is the time between the symptom onset of the primary case and the symptom onset of the secondary case, the PMF of which is represented by $\{\omega_u\}_{u=1}^{u_{max}}$, where $u_{max}$ is the assumed maximum possible serial interval. The term $C_{t-u} \omega_u$ in the summation represents the expected contribution to current reported cases by reported cases from $u$ time steps ago. This relationship is visualised in Figure \ref{fig:background-renewalmodel}.

In a simple example, $R_t$ follows a Gaussian random walk (the state-space transition model), and daily reported cases have an expected value defined by equation (\ref{eq:renewalmodel}) (the observation model). If a Poisson observation distribution is used, then this is the model defined by EpiFilter \cite{paragImprovedEstimationTimevarying2021}, which we also adopt in example model 1.

A more realistic model might assume that underlying infection incidence follows the renewal model: equation \ref{eq:renewalmodel} is modified by replacing observed $C_t$ with unobserved $I_t$, and the serial interval distribution with the generation time distribution. Reported cases are then treated as noisy observations of infection incidence. In this setup, the renewal model now defines part of the state-space transition model, and the observation process is modelled separately. Explicitly modelling infection-to-infection, instead of reported case-to-reported case, has the added benefit of allowing for potentially negative serial intervals, which would otherwise be difficult to account for, especially when fitting models to paired data with observed negative intervals. This is the approach adopted in example models 2 and 3. Crucially, SMC methods can be used in both cases, whereas EpiFilter is limited to the case where the renewal model defines the observation distribution.

The discrete-time nature of our model necessitates the specification of an appropriate time-step duration. The key constraint can be seen in Equation \ref{eq:renewalmodel}, which assumes that primary cases (or infections if we model the underlying infection incidence) cannot produce secondary cases (secondary infections) on the same day they were reported (infected). The discretisation should be fine enough that the probability of same-day transmission is negligible \cite{ogi-gittinsSimulationbasedApproachEstimating2024}. For COVID-19, daily time steps were popular, as this coincided with the frequency of reported case data.

In some scenarios, data may be reported on a longer time scale than infections occur. For example, weekly reporting is common in influenza surveillance, while typical influenza generation times are measured in days \cite{ogi-gittinsSimulationbasedInferenceTimedependent2025a}. Published literature accounts for this using simulation-based methods \cite{ogi-gittinsSimulationbasedApproachEstimating2024, ogi-gittinsSimulationbasedInferenceTimedependent2025a}, expectation-maximisation \cite{nashEstimatingEpidemicReproduction2023}, or Gaussian processes \cite{abbottEstimatingTimevaryingReproduction2020} to infer infection incidence at a finer time-scale before applying the renewal model. The SMC approach introduced in this primer does not assume data are observed on every time step, allowing an appropriate discretisation to be chosen to suit the pathogen being modelled, independently of how frequently data are reported. This is examined in a practical example in Section \ref{sec:temporalmodels}. We also demonstrate this on an influenza dataset \href{https://nicsteyn2.github.io/SMCforRt/models_tempagg.html}{online}.

\section{Sequential Monte Carlo methods}

We introduce two SMC-type algorithms for fitting the renewal model to data. The first is a bootstrap filter, which takes observed data $y_{1:T}$, parameter vector $\theta$, and initial state distribution $P(X_0)$ as inputs, and targets the posterior distribution $P(X_t | y_{1:T}, \theta)$ of the hidden states $X_t$ at each time step $t$. For example, this might be the posterior distributions of $R_t$ (at each $t$) given observed cases $C_{1:T}$ and some smoothing parameter $\eta$. The second algorithm is PMMH, which replaces the fixed $\theta$ with a prior distribution $P(\theta)$ and targets the corresponding posterior distribution $P(\theta|y_{1:T})$.

\begin{figure}[h!]
    \centering
    \includegraphics[width=\textwidth]{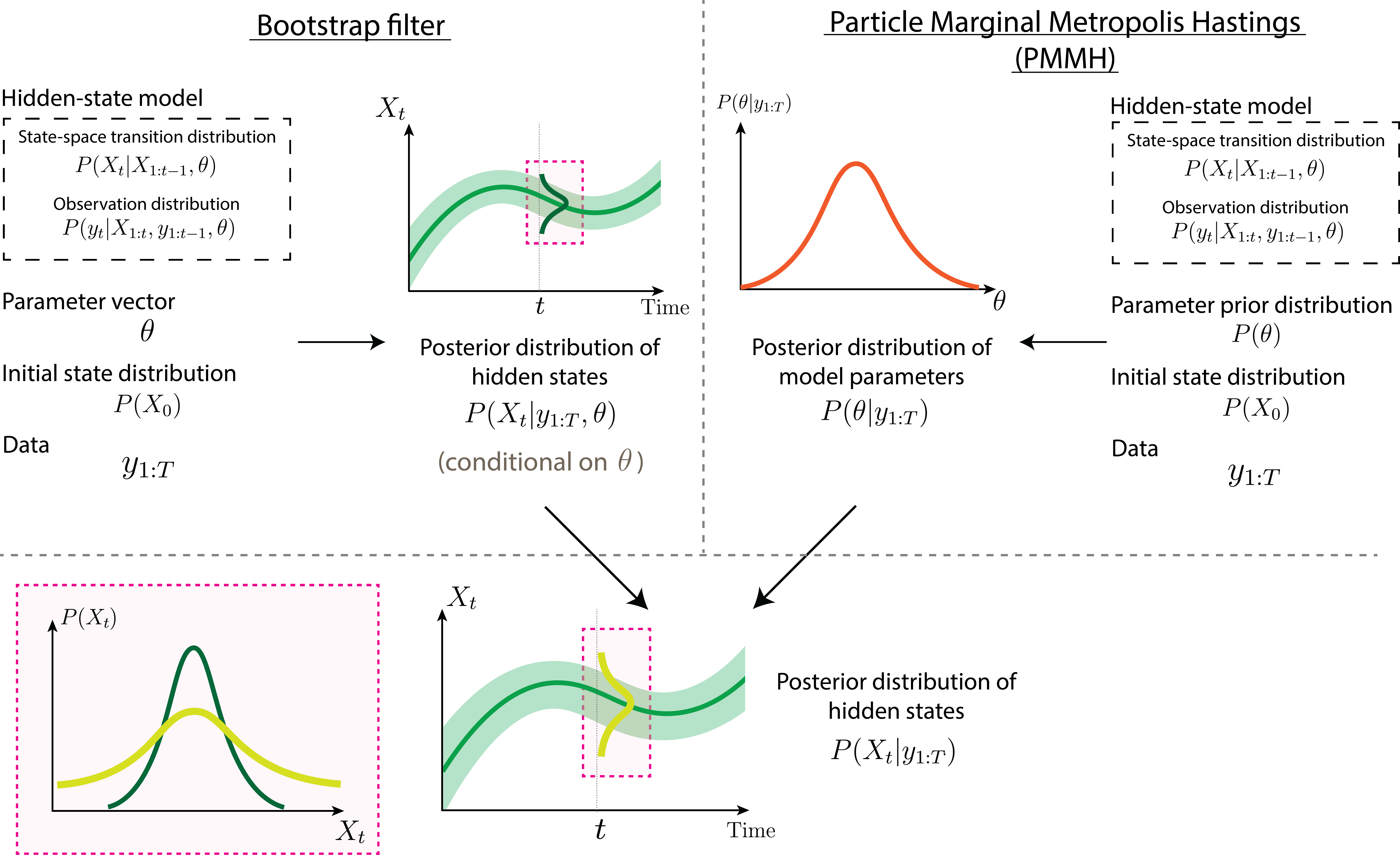}
    \caption{An overview of the algorithms presented in this primer. The bootstrap filter (Algorithm \ref{alg:smc}) and PMMH (Algorithm \ref{alg:pmmh}) both take a hidden-state model, an initial state distribution, and data as inputs. The bootstrap filter additionally requires the specification of parameter vector $\theta$ and returns samples from the posterior distribution of the hidden states at each time step, conditional on $\theta$, shown here as central estimates and credible intervals over time (green curves). PMMH instead requires a prior distribution for $\theta$ and returns samples from the posterior distribution of $\theta$, shown here as the orange curve. These two algorithms can be combined (by running the bootstrap filter at multiple samples of $\theta$ from PMMH) to find the posterior distribution of the hidden states, after accounting for uncertainty about $\theta$ (Algorithm \ref{alg:marginalsmoothing}). The pink inlay shows the posterior distribution of $X_t$ at time $t$ before and after marginalisation, illustrating that uncertainty about $X_t$ is generally greater after accounting for uncertainty about $\theta$. When the bootstrap filter is run with fixed-lag resampling, an additional hyperparameter $L$ is required as an input.}
    \label{fig:algorithms}
\end{figure}

These two algorithms can be used together to estimate the posterior distribution of the hidden states after marginalising over $\theta$, denoted $P(X_t|y_{1:T})$. In the example above, this is the posterior distribution of $R_t$ given $C_{1:T}$, accounting for uncertainty about $\eta$. All three of these posterior distributions can be useful, depending on the model and its intended purpose. An overview of these algorithms is provided in Figure \ref{fig:algorithms}.

SMC methods are particularly useful in epidemiology as they allow for simulation-based inference. The bootstrap filter only requires sampling from the state-space transition distribution, rather than evaluating its density. This makes it possible to fit complicated models where the likelihood is only defined implicitly by a simulator. We still need to be able to evaluate the observation model, but this is often much simpler.

When estimating the value of hidden states, it is common to distinguish between the \textit{filtering} and \textit{smoothing} distributions. The former is the posterior distribution of $X_t$ given data up to time $t$: $P(X_t | y_{1:t}, \theta)$, while the latter is the posterior distribution of $X_t$ given both past and future data: $P(X_t | y_{1:T}, \theta)$ - note the different subscripts on $y$. The uncertainty associated with the smoothing distribution is generally lower than that associated with the filtering distribution as additional data are leveraged, although we are restricted to the filtering distribution in real-time analysis. The bootstrap filter presented below can be used to sample from either distribution.

\subsection{The bootstrap filter}

The goal of the bootstrap filter \cite{gordonNovelApproachNonlinear1993} is to generate samples from the filtering and/or smoothing distributions, conditional on the observed data $y_{1:T}$ and parameter vector $\theta$. This is a similar goal to EpiFilter, for example, which targets the posterior distribution of $R_t$ given observed cases $C_{1:T}$ and parameter $\eta$ \cite{paragImprovedEstimationTimevarying2021}. The bootstrap filter is one of the simplest particle filtering methods, although more advanced methods exist and may be preferable in some settings \cite{chopinIntroductionSequentialMonte2020}. The similarity to EpiFilter, which produces analytical solutions, offers a convenient way to check a bootstrap filter implementation: fitting the same model to the same data allows outputs to be compared directly.

\begin{figure}[h!]
    \centering
    \includegraphics[width=\textwidth]{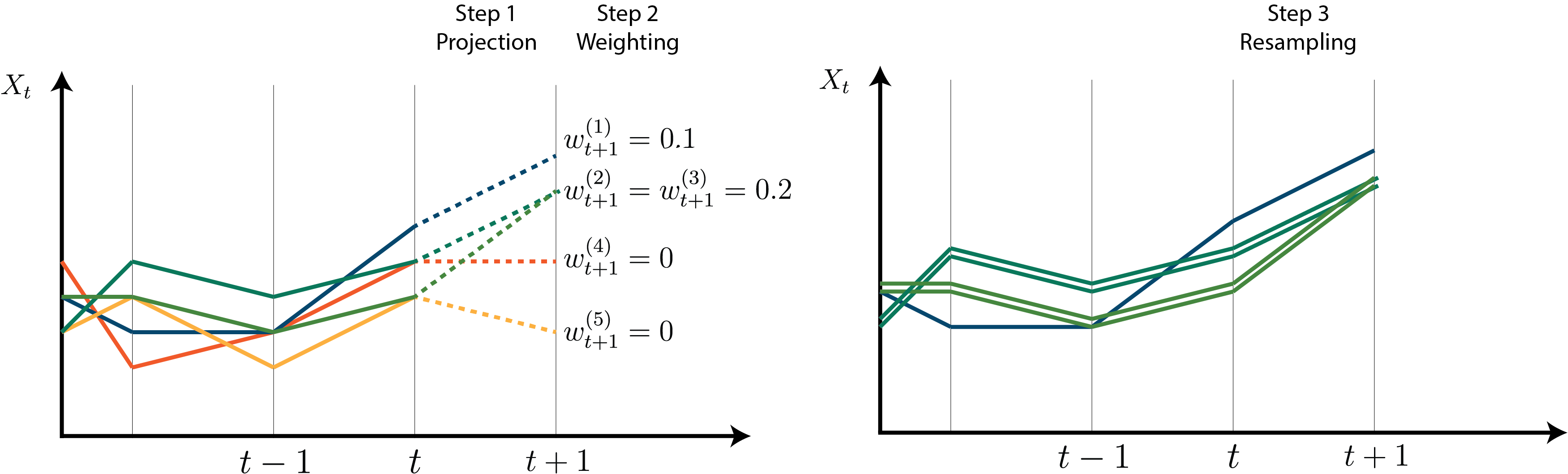}
    \caption{A single step of the bootstrap filter with resampling. Particles are projected forward by sampling according to the state-space transition distribution, weighted by the observation distribution, and resampled to form a new set of equally-weighted particles. In this example, the orange and yellow trajectories receive zero weight, so do not feature in the right-hand panel. The remaining trajectories feature in the right-hand panel with relative frequency proportional to their weight.}
    \label{fig:diag-smc}
\end{figure}

The algorithm starts with an initial set of $N$ ``particles'' $\{x_0^{(i)}\}_{i=1}^N$ sampled from an arbitrary initial-state distribution $P(X_0)$. These particles are projected forward by sampling from the state-space transition distribution $\tilde{x_t}^{(i)} \sim P(X_t | x_{0:t-1}, \theta)$, representing equally likely one-step-ahead projections of the hidden states before any data are observed. The projected particles are then weighted by the observation model $w_t^{(i)} = P(y_t | \tilde{x_t}^{(i)}, y_{1:t-1}, \theta)$, quantifying how likely each projected particle is once the data at time step $t$ are observed. The particles are then resampled with replacement according to these weights, resulting in a new set of particles $\{x_t^{(i)}\}_{i=1}^{N}$ that represent equally-weighted samples from the posterior distribution at time step $t$. This process is described in Figure \ref{fig:diag-smc}. Repeating for $t = 1, \ldots, T$ completes the algorithm, described formally in Algorithm \ref{alg:smc}. Further intuition in the language of sequential importance sampling is provided \href{https://nicsteyn2.github.io/SMCforRt/smc-bootstrap.html}{online}. Readers familiar with alternative SMC methods may expect all weights to be reset to $1/N$ following resampling. This is unnecessary as we resample at every step, thus we always treat the resampled particles as equally-weighted samples from the target posterior distribution and past weights do not feature in calculations at future time steps (that is, there is nothing to ``reset''). Furthermore, the unnormalised weights $w_t^{(i)}$ are later used when estimating the likelihood in Algorithm \ref{alg:pmmh}, so it is helpful to store these as initially calculated.

If only the particles $x_t^{(i)}$ at time step $t$ are resampled, then equally-weighted samples from the filtering distributions $P(X_t | y_{1:t}, \theta)$ are returned (for each $t$). If all particle histories $x_{1:t}^{(i)}$ are resampled at each time step (as in Figure \ref{fig:diag-smc}), then equally-weighted samples from the smoothing distributions $P(X_t | y_{1:T}, \theta)$ are returned. Intuitively, this is because resampling effectively re-weights past particles according to current data.

When $T$ is large, resampling entire particle histories leads to particle degeneracy \cite{sarkkaBayesianFilteringSmoothing2013}, where few unique particles remain at early time steps. This is visible in Figure \ref{fig:diag-smc} as a decreasing number of unique particle trajectories after resampling, which results in poor approximations of the smoothing distribution. To mitigate particle degeneracy, we use fixed-lag resampling \cite{chopinIntroductionSequentialMonte2020}, resampling the past $L$ time steps $x_{t-L:t}^{(i)}$ rather than the full state histories, which yields an approximation to the smoothing distribution: $P(X_t | y_{1:t+L}, \theta)$. If $X_t$ is conditionally independent of data more than $L$ steps in the future, given $y_{1:t+L-1}$, then this approximation is exact. In renewal models, $L$ should exceed the maximum serial interval $u_{max}$, and, if reporting delays are present, $L$ should also exceed the maximum reporting delay.

Thus far, we have assumed we are targeting the marginal posterior distribution at time $t$. It may also be useful to consider the joint distribution of the hidden states over multiple time steps, $P(X_{s:t} | y_{1:T}, \theta)$ for some $s < t$. This is useful if we are estimating the timing of epidemic peaks (such as peak infections or peak $R_t$), for example. If full state histories are resampled at each time step, then the resulting trajectories $x_{1:T}^{(i)}$ are jointly distributed according to $P(X_{1:T} | y_{1:T}, \theta)$. If fixed-lag resampling is used, then the final $L$ time steps of each trajectory $x_{T-L:T}^{(i)}$ are jointly distributed according to $P(X_{T-L:T} | y_{1:T}, \theta)$, where $T$ is the time step at which the algorithm is halted.

\begin{algorithm}
    \caption{\underline{The fixed-lag bootstrap filter.} Statements indexed by $i$ should be taken to mean for $i$ in $\{1, 2, \ldots, N\}$. At time step $t$, $\tilde{x}_{t-L:t}^{(i)}$ is taken to mean the aggregation of particle history $x_{t-L:t-1}^{(i)}$ and newly projected particles $\tilde{x}_{t}^{(i)}$.}
    \label{alg:smc}
    \begin{algorithmic}[1]
    \State \textbf{Input:} Number of particles $N$, fixed lag $L$, parameter vector $\theta$, data $y_{1:T}$, state-space transition distribution $P(X_t | X_{1:t-1}, \theta)$, observation distribution $P(y_t | X_{1:t}, y_{1:t-1}, \theta)$, and initial state-space distribution $P(X_0)$
    \State \textbf{Sample}  $x_{0}^{(i)} \sim P(X_0)$
    \For{$t = 1$ to $T$}
    \State $\tilde{x}_{t}^{(i)} \sim P(X_t | x_{0:t-1}^{(i)}, \theta)$ \Comment{Sample from the state-space transition distribution}
    \State $w_{t}^{(i)} \gets P(y_t | \tilde{x}_{t-L:t}^{(i)}, y_{1:t-1}, \theta)$ \Comment{Calculate the observation weights}
    \State $x_{t-L:t}^{(i)} \sim \text{Multinomial}\left(\{\tilde{x}_{t-L:t}^{(j)}\}_{j=1}^N, \left\{ \frac{w_{t}^{(j)}}{\sum_{k=1}^N w_{t}^{(k)}} \right\}_{j=1}^N\right)$ \Comment{Resample the particles}
    \EndFor
    \State \textbf{Return:} Particle values $x_{t}^{(i)}$ and observation weights $w_{t}^{(i)}$ for $i = 1, \ldots, N$ and $t = 1, \ldots T$.
    \end{algorithmic}
\end{algorithm}

\subsubsection{Practical considerations}

The choice of fixed lag $L$ depends on the modelled maximum serial interval (or generation time) $u_{max}$. In many cases, such as in the COVID-19 examples considered, $u_{max}$ (and thus $L$) can be chosen to be sufficiently small to ensure algorithmic efficiency without introducing noticeable bias. Some diseases, such as tuberculosis \cite{vinkSerialIntervalsRespiratory2014}, feature serial intervals and generation times that are measured in weeks, months, or even years.  While this poses a computational problem when using a daily discretisation (requiring a very large $L$), the longer serial intervals/generation times allow for the pathogen to be modelled on a weekly (or even monthly) discretisation, thus $L$ can be kept small. The impact of $L$ can be checked by refitting the model multiple times at different values of $L$ and comparing the outputs. This is demonstrated \href{https://nicsteyn2.github.io/SMCforRt/smc-combining.html}{online}, where we find that reasonable choices of $L$ have very little impact on an example model.

A greater number of particles $N$ improves the accuracy of the posterior distribution approximation but increases computation time. The appropriate value of $N$ depends on the complexity of the model, how well the model fits the data, and the purpose of the bootstrap filter, and thus is difficult to determine a priori. In practice, we find that $N = 100{,}000$ is sufficient (i.e., produces stable central estimates and credible intervals) for all models considered in this primer. This can be decreased (typically to around $N = 1{,}000$) when the algorithm is used for likelihood estimation (see Section \ref{sec:pmmh} for further discussion) or when the algorithm is used to find the marginal smoothing posterior distribution (Section \ref{sec:robusthiddenstate}).

The choice of $N$ for a specific model can be guided by comparing hidden-state estimates across multiple runs: if the estimates are stable (i.e., they do not noticeably change between runs), then $N$ is likely sufficiently large. If a programmatic heuristic is required, the particle weights (line 5 of Algorithm \ref{alg:smc}) can be used to calculate the effective sample size\cite{elviraRethinkingEffectiveSample2022}, which is inversely proportional to the variance of the estimate of the posterior mean and credible interval width. The calculation of effective sample size is demonstrated on an example model \href{https://nicsteyn2.github.io/SMCforRt/smc-bootstrap.html#choosing-the-number-of-particles}{online}.  Further validation can be performed by simulating data from the model and checking that the model recovers the simulated truth, or by fitting a similar model with a known analytical solution (such as EpiFilter) and comparing results to the known ground truth. Theoretical convergence results \cite{doucetTutorialParticleFiltering2011} developed for Markovian state-space models can be applied by augmenting past states into $X_t$ and considering the model on the expanded state space.

In Algorithm \ref{alg:smc}, for simplicity, we use the state-space transition distribution as the proposal distribution. This is known to be suboptimal, particularly because it does not use information about the observed data $y_t$ when proposing new particles. More complex proposal distributions can be employed and the weights adjusted to account for the fact that the proposal distribution is not the true state-space transition distribution. One such example is the auxiliary particle filter \cite{pittFilteringSimulationAuxiliary1999}, although we find this unnecessary for the models fit in this primer.

Due to fixed-lag resampling, estimates are usually robust to reasonable choices of the initial state distribution $P(X_0)$. In the absence of clear prior information, the initial state distribution should allow for any plausible value of $X_0$. The effect of the initial state distribution can be examined through sensitivity analysis.

A large proportion of SMC literature focuses on hidden Markov Models. In this primer we apply SMC methods to non-Markovian models (i.e. those where $X_t$ depends on $X_{1:t-1}$ rather than just $X_{t-1}$). To ensure past particle trajectories are consistent, fixed-lag resampling must be used when fitting non-Markovian models, even when only targeting the filtering posterior distribution. In these cases, $L$ should be chosen such that $X_t$ does not depend on hidden-state values prior to $t-L$. For example, when the state-space transition model includes the renewal model, $L$ should be chosen greater than the maximum serial interval.

\subsection{Particle marginal Metropolis-Hastings}
\label{sec:pmmh}

The goal of the PMMH algorithm \cite{andrieuParticleMarkovChain2010} is to find the posterior distribution of parameter vector $\theta$, given observed data $y_{1:T}$. We do this by employing a Metropolis-Hastings algorithm \cite{gelmanBayesianDataAnalysis2014}, while using the bootstrap filter (Algorithm \ref{alg:smc}) to estimate the likelihood of the proposed parameters.

The log-likelihood estimator is given by the logarithm of the geometric mean of the unnormalised particle weights $\bar{w}_t$ from the bootstrap filter (Algorithm \ref{alg:smc}) at each time step \cite{endoIntroductionParticleMarkovchain2019, hurzelerApproximatingMaximisingLikelihood2001}:

\begin{equation}
    \hat{\ell}(\theta|y_{1:T}) = \frac{1}{T} \sum_{t=1}^T \log \bar{w}_t \label{eqn:likelihoodest}
\end{equation}

\noindent Equation \ref{eqn:likelihoodest} follows immediately from the predictive decomposition of the likelihood:

\begin{equation}
    \ell(\theta|y_{1:T}) = \log P(y_{1:T} | \theta) = \sum_{t=1}^T \log P(y_t | y_{1:t-1}, \theta) \label{eqn:predictivedecomp}
\end{equation}

\noindent where a Monte Carlo estimator of the one-step-ahead predictive likelihood is given by the average of the unnormalised weights of the projected particles, denoted $\bar{w}_t$:

\begin{equation}
    \begin{aligned}
    P(y_t | y_{1:t-1}, \theta) &= E_{X_{1:t}|y_{1:t-1}}\left[P(y_t | X_{1:t}, y_{1:t-1}, \theta) \right]\\
    &\approx \frac{1}{N} \sum_{i=1}^N P\left(y_t | \tilde{x}_{t}^{(i)}, y_{1:t-1}, \theta\right)\\
    &= \bar{w}_t
    \end{aligned}
\end{equation}

\noindent Parameter inference then proceeds by using $\hat{\ell}(\theta|y_{1:T})$ in the acceptance probability of an otherwise standard Metropolis-Hastings algorithm (Algorithm \ref{alg:pmmh}) .

\begin{algorithm}
    \caption{\underline{Particle marginal Metropolis Hastings (PMMH).} The bootstrap filter $\mathcal{M}$ written here accepts proposed parameters $\theta'$ and returns unnormalised weights $\{w_t^{(i)}\}$ for $i = 1, \ldots, N$ and $t = 1, \ldots, T$. The average of all weights at time $t$ is denoted $\bar{w}_t$. The data $y_{1:T}$ and other bootstrap filter options are assumed to be included in $\mathcal{M}$. }
    \label{alg:pmmh}
    \begin{algorithmic}[1]
    \State \textbf{Input:} Number of parameter samples $M$, bootstrap filter $\mathcal{M}$, parameter prior distribution $P(\theta)$, parameter proposal distribution $q(\theta'|\theta)$.
    \State \textbf{Initialise:} $\theta_1 \sim P(\theta)$
    \For{$j = 2$ to $M$}
    \State $\theta' \sim q(\theta' | \theta_{j-1})$ \Comment{Sample proposed parameters}
    \State $\{w_t^{(i)}\}_{i = 1, \ldots, N, \ t = 1, \ldots, T} \gets \mathcal{M}(\theta')$ \Comment{Run the bootstrap filter (Algorithm \ref{alg:smc}) at $\theta'$}
    \State $\hat{\ell}(\theta') = \gets \frac{1}{T} \sum_{t=1}^T \log \bar{w}_t$ \Comment{Calculate log-likelihood estimate (equation \ref{eqn:likelihoodest})}
    \State $\alpha \gets \min\left(\frac{\exp \hat{\ell}(\theta')}{\exp \hat{\ell}(\theta_{j-1})}\frac{P(\theta')}{P(\theta_{j-1})} \frac{q(\theta_{j-1}|\theta')}{q(\theta'|\theta_{j-1})}, 1\right)$ \Comment{Calculate acceptance probability}
    \If{$U \sim \text{Uniform}(0, 1) < \alpha$}
    \State $\theta_j \gets \theta'$ \Comment{Accept the proposal}
    \Else
    \State $\theta_j \gets \theta_{j-1}$ \Comment{Reject the proposal}
    \EndIf
    \EndFor
    \State \textbf{Return:} Sampled parameter values $\{\theta_j\}_{j=1}^M$.
    \end{algorithmic}
\end{algorithm}

\subsubsection{Practical considerations}

To enable the calculation of convergence diagnostics, multiple chains (we find 4 works well) of the PMMH algorithm should be run. We also encourage the standard practice of discarding an initial portion of each chain as a burn-in period, ensuring the retained samples are not influenced by initial values outside the stationary distribution of the Markov chain. Diagnostics used in the examples below are the Gelman-Rubin statistic $\hat{R}$ \cite{gelmanInferenceIterativeSimulation1992} and effective sample size (ESS) \cite{gelmanBayesianDataAnalysis2014}, which are reported by standard Markov chain Monte Carlo (MCMC) analysis software. We use the MCMCChains.jl package \cite{ge2018t} in our examples.

The standard Metropolis-Hastings algorithm uses exact evaluations of the model log-likelihood $\ell(\theta)$, whereas Algorithm \ref{alg:smc} admits a stochastic estimator $\hat{\ell}(\theta)$. The standard deviation of this estimator is a function of the number of particles $N$ used in the bootstrap filter. It has been shown that the optimal standard deviation of $\hat{\ell}(\theta)$ is 1.2-1.3 \cite{kantasParticleMethodsParameter2015, doucetEfficientImplementationMarkov2015}, which can be used to guide the choice of $N$. The standard deviation of the log-likelihood estimate can itself be estimated by refitting the model multiple times, we demonstrate this \href{https://nicsteyn2.github.io/SMCforRt/smc-pmmh.html#variance-of-log-likelihood-estimates}{online}. For all examples in this primer, we use $N = 1,000$ particles when estimating model likelihoods. We also note that, while larger standard deviations of $\hat{\ell}(\theta)$ result in slower convergence, model outputs are still valid. 

The efficiency of the PMMH algorithm depends on the parameter proposal distribution $q(\theta' | \theta)$. We find that the heuristic multivariate normal proposal distribution with covariance matrix $\Sigma = (2.38^2/d) \hat{\Sigma}$, where $\hat{\Sigma}$ is the sample covariance matrix of previous samples and $d$ is the number of parameters being fit, generally performs well \cite{andrieuTutorialAdaptiveMCMC2008}.

In the examples presented in this primer, we begin the PMMH algorithm with a diagonal covariance matrix and update it every 100 iterations until it stabilises (once $|\Sigma|$ changes by less than 20\% between iterations), and then begin the primary sampling. Primary sampling is also performed in chunks of 100 samples, with the $\hat{R}$ and ESS calculated at the end of each chunk. We stop the algorithm when $\hat{R} < 1.05$ and ESS $> 100$ for all parameters. Complete details of these procedures are provided in the model files \href{https://github.com/nicsteyn2/SMCforRt/tree/main/paper}{in the GitHub repository}.

\section{General framework}

Thus far we have developed an algorithm for estimating the posterior distribution of hidden states $X_t$ given observed data $y_{1:T}$ and parameter vector $\theta$, and an algorithm for estimating the posterior distribution of $\theta$ given $y_{1:T}$. In this section we demonstrate how these two algorithms can be used to perform inference and generate projections.

\subsection{Robust hidden-state inference}
\label{sec:robusthiddenstate}

We can couple algorithms \ref{alg:smc} and \ref{alg:pmmh} to estimate the marginal smoothing distribution $P(X_t | y_{1:T})$, representing our beliefs about $X_t$ after accounting for uncertainty about $\theta$. Intuitively, we use the bootstrap filter to fit the model at multiple parameter samples $\theta^{(i)} \sim P(\theta|y_{1:T})$ and average the results:

\begin{equation}
    \begin{aligned}
    P(X_t | y_{1:T}) &= E_{\theta|y_{1:T}}\left[P(X_t|y_{1:T}, \theta)\right]\\
    &\approx \frac{1}{N} \sum_{j=1}^N P(X_t | y_{1:T}, \theta^{(j)}) \quad \text{where} \quad \theta^{(j)} \sim P(\theta | y_{1:T})
    \end{aligned}
\end{equation}

In practice, this is achieved by sampling $N_\theta$ parameter samples from the output of the PMMH algorithm (Algorithm \ref{alg:pmmh}), and running the bootstrap filter (Algorithm \ref{alg:smc}) at each of these samples. $N$ particles are used in each bootstrap filter, resulting in a total of $N_\theta N$ particles approximating the marginal smoothing posterior. In the examples below, we use $N = 1{,}000$ and $N_\theta = 100$. These values are chosen to ensure that (a) a sufficiently diverse set of parameter samples is used (selection of $N_\theta$), and (b) the posterior distribution of the hidden states is well approximated (selection of $N_\theta N$). We assess these criteria by checking that the marginal smoothing distribution is stable across multiple runs of the algorithm.

\begin{algorithm}
    \caption{\underline{Marginal smoothing distribution sampler} The bootstrap filter $\mathcal{M}$ written here accepts a parameter vector $\theta$ and returns a matrix of hidden state values. Note that inputs must satisfy $N_p = N_\theta N$.}
    \label{alg:marginalsmoothing}
    \begin{algorithmic}[1]
    \State \textbf{Input:} Parameter samples from PMMH $\{\theta_j\}_{j=1}^M$, number of unique parameter samples $N_\theta$, per-filter posterior samples $N$, target posterior samples $N_p$, and bootstrap filter $\mathcal{M}$.
    \State \textbf{Initialise:} Pre-allocate matrix $X$ of size $N_p \times T$.
    \For{$i = 1, \ldots, N_\theta$}
    \State $\theta' \sim \{\theta_j\}_{j=1}^{M}$ \Comment{Sample a parameter vector $\theta'$}
    \State $\mathcal{I} \gets \{(i - 1) N + 1, \ldots, i N\}$ \Comment{Specify the indices of the $i$th block of $X$}
    \State $X_{\mathcal{I}, 1:T} \gets \mathcal{M}(\theta')$ \Comment{Run the bootstrap filter (Algorithm \ref{alg:smc}) at $\theta'$}
    \EndFor
    \end{algorithmic}
\end{algorithm}

\subsection{Predictions}

Thus far we have focused on model inference, both for hidden states $X_t$ and parameter vector $\theta$. We may also be interested in interpolating missing data $Y_t$, extrapolating future data $Y_{T+k}$, or projecting future hidden states $X_{T+k}$.

\subsubsection{Predictive posterior distribution of observed data}

Various predictive posterior distributions can be targeted. For example, the one-step-ahead predictive posterior distribution $P(Y_t|y_{1:t-1}, \theta)$ can be targeted within the bootstrap filter by sampling from the observation distribution given the projected particles. That is, sampling $y_t^{(i)} \sim P(Y_t | \tilde{x}_{1:t}^{(i)}, \theta)$, where $\tilde{x}_{1:t}^{(i)}$ are the projected particles as defined on line 4 of Algorithm \ref{alg:smc}.

Alternatively, the smoothing posterior predictive distribution $P(Y_t|y_{1:T}, \theta)$ can be targeted within the bootstrap filter by sampling from the observation distribution given the projected particles at each time step $t$ as above (setting $\tilde{y}_t^{(i)} \sim P(Y_t | \tilde{x}_{1:t}^{(i)}, \theta)$). Defining $\tilde{y}_{t-L:t}^{(i)}$ as the aggregation of the newly sampled $\tilde{y}_t^{(i)}$ and previously stored $y_{t-L:t-1}^{(i)}$, and resampling $y_{t-L:t}^{(i)}$ from $\tilde{y}_{t-L:t}^{(i)}$ alongside $x_{t-L:t}^{(i)}$ in step 6 of the Algorithm \ref{alg:smc} results in a set of particles $\{y_t^{(i)}\}$ that can be viewed as unweighted samples from the smoothing posterior predictive distribution. This can be clearly seen if we consider the predictive variable as part of an extended hidden state space. To ensure consistency between the weights calculated in step 5 of Algorithm \ref{alg:smc} and the predictive particles, it is important to use the same indices when resampling both $x_t^{(i)}$ and $y_t^{(i)}$. Computationally this is equivalent to sampling some indices from $\text{Multinomial}\left(\{1, 2, \ldots, N\}, \left\{ \frac{w_{t}^{(j)}}{\sum_{k=1}^N w_{t}^{(k)}} \right\}_{j=1}^N \right)$, and then using these to resample both the hidden states and predictive variable. The parameter vector $\theta$ can be marginalised out by including and storing the relevant samples in Algorithm \ref{alg:marginalsmoothing}.

\subsubsection{Missing data}

Missing data can be handled in a variety of ways depending on the type of missingness (missing at random, missing not at random, etc.\cite{rubinInferenceMissingData1976}). Where data are missing by design (for example, wastewater sampling data may only be collected on certain days \cite{watsonJointlyEstimatingEpidemiological2024}), the observation distribution becomes $P(y_t \text{ is missing} | X_{1:t}, y_{1:t-1}, \theta) = 1$, irrespective of the values of the hidden states. In practice, we also skip the resampling step in the bootstrap filter on such days, as all hidden states are equally likely. More complex models, where the relationship between hidden states and missingness is explicitly modelled, are also possible. Prediction of missing data can be made by sampling from the observation model, conditional on the value of the hidden states (another type of predictive posterior distribution).

\subsubsection{Projections}

The projection of future hidden states (such as future $R_t$) is handled similarly, effectively treating future data as missing by design, and projecting $X_T$ forward by repeatedly sampling from the state-space transition distribution, generating projections of hidden states $X_{T+k}$. Observed data $Y_{T+k}$ can then be projected made by sampling from the observation distribution conditional on the projected hidden states. We demonstrate this in example (3) below.

\subsubsection{Probability of elimination}

Elimination of an infectious disease is generally defined as ``the reduction to zero incidence of a certain pathogen in a given area'' \cite{klepacSixChallengesEradication2015}. For the purposes of this primer, we say that elimination has occurred at time $t$ if no new infections occur within $[t, t+28]$, an appropriate window for the SARS-CoV-2 examples we consider, although a variety of other mathematical statements matching this definition are possible \cite{plankEstimationEndofoutbreakProbabilities2025, paragDecipheringEarlywarningSignals2021}. This turns the problem of estimating the probability of elimination into a projection problem: we project the number of new infections in the next 28 days, and the proportion of particle trajectories that contain zero new infections provides an estimate of the probability of elimination. We demonstrate this in example (2) below.

\subsection{Model evaluation and selection}
\label{sec:modelselection}

Model evaluation and selection are crucial components of any modelling exercise. We outline three metrics: root mean square error (RMSE) of the posterior predictive distribution, the coverage of posterior predictive credible intervals, and the continuous ranked probability score (CRPS) \cite{gneitingStrictlyProperScoring2007} of the posterior predictive distribution.

The RMSE of the posterior predictive distribution quantifies the average discrepancy between observed data and the model's predictions. It is calculated as:
\begin{equation}
    \text{RMSE} = \sqrt{\frac{1}{T} \sum_{t=1}^T (y_t - \hat{y}_t)^2}
\end{equation}
where $\hat{y}_t = E[Y_t | y_{1:T}]$ is the expected value of the posterior predictive distribution at time $t$. Smaller values indicate a better fit to the data.

The coverage of posterior predictive credible intervals measures the proportion of observed data that falls within the model's credible intervals. It is calculated as:
\begin{equation}
    \text{Coverage}_\alpha = \frac{1}{T} \sum_{t=1}^T \mathbb{I}(y_t \in \text{CI}_{\alpha, t})
\end{equation}
where $\text{CI}_{\alpha, t}$ is the $(1-\alpha)$-level credible interval of the posterior predictive distribution at time $t$. A well-calibrated model yields credible intervals whose empirical coverage is close to the nominal level on average, for all values of $\alpha$. For example, if $1-\alpha = 0.95$, then the posterior predictive credible intervals should contain the observed data approximately 95\% of the time.

Finally, the CRPS of the posterior predictive distribution quantifies the average discrepancy between the predictive cumulative distribution function (CDF) and the empirical CDF. It is defined as:
\begin{equation}
    \text{CRPS} = \frac{1}{T} \sum_{t=1}^T \left(\int_{-\infty}^\infty (P(Y_t \leq y|y_{1:T}) - \mathbb{I}(y_t \leq y))^2 dy\right)
\end{equation}
Smaller values of CRPS indicate a posterior CDF that more closely matches the empirical CDF of the observed data.

In practice, we use a particle approximation to the expectation-definition of the CRPS \cite{jordanEvaluatingProbabilisticForecasts2019}:
\begin{equation}
    \begin{aligned}
        \text{CRPS} &= \frac{1}{T} \sum_{t=1}^T \left[ \frac{1}{2} E|Y_t - Y_t'| - E|Y_t - y| \right] \quad \text{where} \quad Y_t, Y_t' \sim P(Y_t | y_{1:T})\\
        &\approx \frac{1}{T} \sum_{t=1}^T \left[ \frac{1}{N} \sum_{j=1}^N |Y_t^{(j)} - y_t| - \frac{1}{2} \frac{1}{N^2} \sum_{j=1}^N \sum_{k=1}^N |Y_t^{(j)} - Y_t^{(k)}| \right] \\
        &= \frac{1}{T} \sum_{t=1}^T \left[ \frac{1}{N} \sum_{j=1}^N |Y_t^{(j)} - y_t| - \frac{1}{N^2} \sum_{j=1}^N (2j - N - 1)\tilde{Y}^{(j)} \right]
    \end{aligned}
\end{equation}
where $Y_t^{(j)}$ are samples from the posterior predictive distribution at time $t$ and $\tilde{Y}^{(j)}$ are the samples sorted in ascending order. Using sorted samples reduces the computational complexity of the CRPS from $\mathcal{O}(N^2)$ (line 2) to $\mathcal{O}(N \log N)$ (line 3) \cite{jordanEvaluatingProbabilisticForecasts2019}. The CRPS is particularly useful in the model selection process, as it provides a principled trade-off between precision (how narrow the credible intervals are) and calibration.

\section{Example: COVID-19 in Aotearoa New Zealand}

We demonstrate these methods by fitting three models to national data from the COVID-19 pandemic in Aotearoa New Zealand \cite{ministryofhealthnzNewZealandCOVID192024}. Two time periods are considered: 26 February 2020 - 4 June 2020 and 1 April 2024 - 9 July 2024. The former is characterised by a single epidemic wave with high-quality case reporting and a large proportion of imported cases, during which elimination of transmission was the primary public health policy aim. The latter is characterised by widespread transmission with a clear day-of-week effect and high levels of reporting noise and bias, with modelling primarily used to inform health-service resource allocation.

\begin{figure}[h!]
    \centering
    \includegraphics[width=\textwidth]{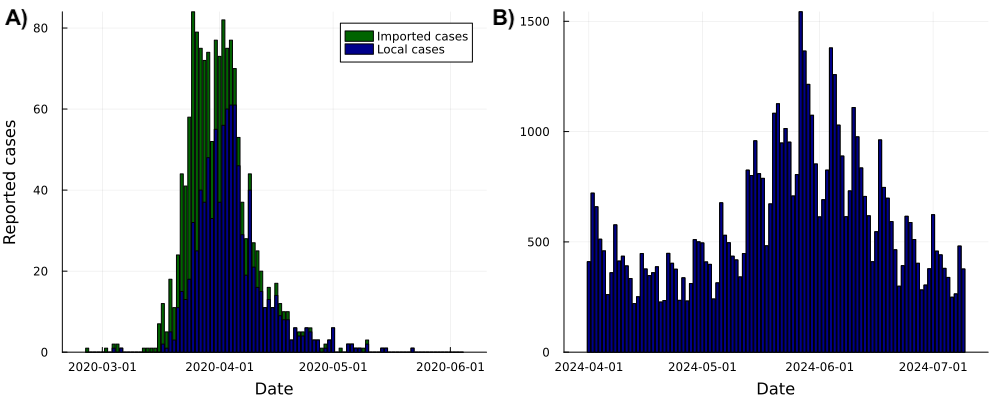}
    \caption{Reported cases from the COVID-19 pandemic in New Zealand, separated by locally-acquired and imported infections, for two time-periods: (A) 26 February 2020 - 4 June 2020 and (B) 1 April 2024 - 9 July 2024 \cite{ministryofhealthnzNewZealandCOVID192024}. The first period is the first wave of COVID-19 in New Zealand, characterised by a single epidemic wave which ended in temporary elimination of local transmission. The second period is characterised by widespread transmission with much higher numbers of reported cases.}
    \label{fig:data}
\end{figure}

\subsection{Model 1: Simple example}
\label{sec:simpleexample}

As a first example, we implement a model very similar to EpiFilter \cite{paragImprovedEstimationTimevarying2021}, assuming that $R_t$ follows a log-normal random walk (the state-space transition model):

\begin{equation}
   \log R_t | \log R_{t-1} \sim \text{Normal}(\log R_{t-1}, \sigma)
\end{equation}

\noindent and observed cases follow the Poisson renewal model (the observation model):

\begin{equation}
    C_t | R_t, C_{1:t-1} \sim \text{Poisson}\left(R_t \sum_{u=1}^{u_{max}} C_{t-u} \omega_u\right)
\end{equation}

In notation from the hidden-state models section, the hidden states are $X_t = R_t$, the observed data are $y_t = C_t$, and the parameter vector is $\theta = \sigma$. We could additionally consider the values of the serial interval PMF $\{\omega_u\}_{u=1}^{u_{max}}$ as model parameters, but for simplicity we treat this is as fixed and part of the model structure, a typical assumption in such models.

The serial interval is assumed to follow a Gamma distribution with mean 6.5 days and standard deviation 4.2 days. For simplicity, this is discretised by evaluating the density at $t = 1, 2, \ldots, T$ days and normalising.

First we use PMMH (Algorithm \ref{alg:pmmh}) to estimate the posterior distribution of $\sigma$. Assuming a uniform prior distribution on $(0, 1)$, we find a posterior mean of 0.24 (95\% Credible Interval (Cr.I.) 0.17, 0.33). This parameter governs the smoothness of $R_t$, though its value is not directly interpretable on its own. Convergence (in this case, $\hat{R} = 1.004$ and ESS $= 171$) was obtained in 200 iterations, taking approximately 10 seconds on a 2021 M1 MacBook Pro.

\begin{figure}[h!]
    \centering
    \includegraphics[width=\textwidth]{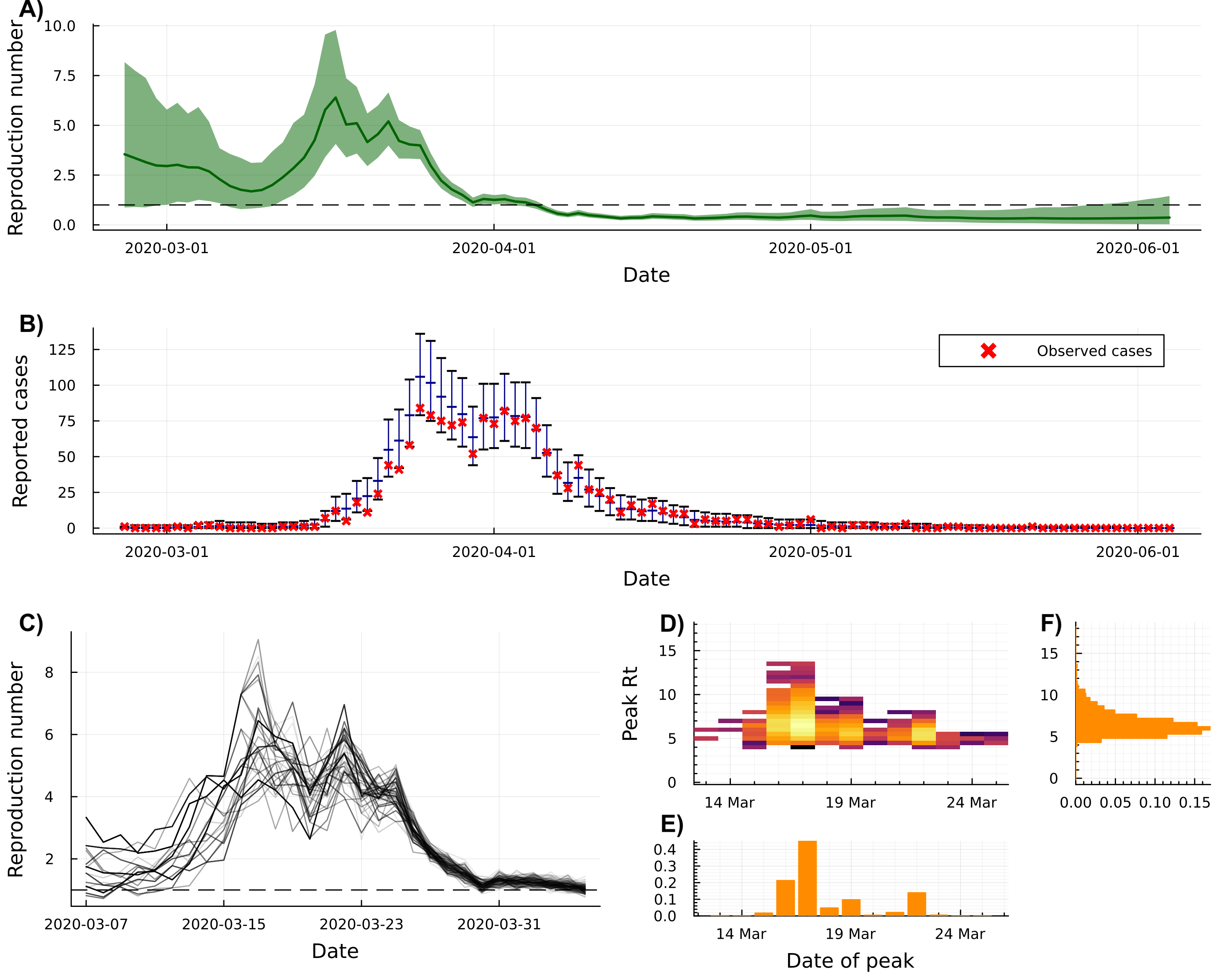}
    \caption{Results from fitting model 1 to data from the COVID-19 pandemic in New Zealand between 26 February 2020 and 4 June 2020. The marginal posterior means and 95\% credible intervals of $R_t$ (panel A) demonstrate high uncertainty in the early stages as well as a high peak estimate of $R_t$ in late-March 2020. The marginal posterior predictive distributions of reported cases (panel B) shows that the model fits the data well, with 95.4\% of non-zero reported daily cases in the dataset falling inside the 95\% posterior predictive credible intervals. Panel (C) shows equally-likely samples from the joint posterior distribution of $R_t$ between 7 March and 5 April 2020. These samples can be used to find the joint posterior distribution of peak $R_t$ and the date of this peak, reported as a log-density heatmap (panel D), where brighter colours represent more likely combinations of these values. The marginal posterior distributions of peak $R_t$ and the date of the peak are shown in panels E and F, respectively. Panels D and E suggests that the peak in $R_t$ (the point at which the SARS-CoV-2 virus was spreading most rapidly) was most likely around 17 March 2020, although a second mode is visible with lower peak $R_t$ on 22 March 2020.}
    \label{fig:simplemodel}
\end{figure}

The focus of this example is on hidden-state inference: the estimation of $R_t$. We produce daily estimates of $R_t$ and corresponding 95\% credible intervals using Algorithm \ref{alg:marginalsmoothing}, effectively averaging over the posterior uncertainty about $\sigma$. These estimates are shown in panel A of Figure \ref{fig:simplemodel}.

We also present the posterior predictive distribution of reported cases in Figure \ref{fig:simplemodel}-B. We assess the calibration of the credible intervals for reported cases by comparing the observed data to the posterior predictive intervals and find that the model fits the data well, with 97.9\% of all daily reported cases and 95.4\% of non-zero daily reported cases in the dataset falling inside the daily 95\% posterior predictive credible intervals.

By halting the SMC algorithm on 5 April 2020 and using $L = 30$ as the resampling lag, we can generate samples from the joint marginal distribution $P(R_{\text{7 March}:\text{5 April}}|C_{\text{26 Feb}:\text{5 April}})$, presented in Figure \ref{fig:simplemodel}-C. These samples allow us to estimate the joint posterior distribution of the peak in transmission (the greatest value of $R_t$) and the timing of this peak. This is reported as a heatmap of the log-posterior density in Figure \ref{fig:simplemodel}-D, where brighter colours represent more likely combinations of these values. The marginal posterior distributions of peak $R_t$ and the date of this peak are shown in panels E and F of Figure \ref{fig:simplemodel}. We estimate that $R_t$ peaked at 6.8 (95\% Cr.I. 4.9, 10.3) on 17 March 2020 (95\% Cr.I. 16 March, 22 March). Estimates of peak $R_t$ and the timing of this peak are useful when evaluating the effect of public health interventions, although we caution that reporting delays should be considered when interpreting these results.

Two distinct modes can be seen in the log-posterior density heatmap: one with a peak on 17 March 2020 and the other with a peak on 22 March 2020. The peak on 17 March is associated with a higher value of 7.3 (5.2, 10.6) (if we condition on this peak date), while a peak on 22 March 2020 is estimated to be lower, at 5.8 (4.8, 7.1). That is, there is evidence for either an early and high peak, or a later and slightly lower peak in transmission.

\subsection{Model 2: Reporting noise, imported cases, and elimination probabilities}
\label{sec:noiseimportselim}

Surveillance of COVID-19 in New Zealand was generally considered to be of high quality, although it can still be desirable to account for certain known biases. Firstly, infection incidence is not directly observed. Instead, we observe reported cases which are subject to reporting noise. Secondly, a large proportion of reported cases in the early phase of the pandemic in New Zealand were imported cases, those infected outside of New Zealand. Failure to account for these may lead the model to overestimate $R_t$, as imported cases are attributed to local transmission.

During this period, elimination of SARS-CoV-2 was an explicit public-health policy goal in New Zealand. Models explicitly estimating the probability of elimination were of key interest to policymakers \cite{hendyMathematicalModellingInform2021}, particularly as repeated days of zero cases were observed. We demonstrate how these methods can be used to estimate the probability of elimination, defined here as the probability that the true infection incidence $I_t$ is zero for the next 28 days (i.e. $I_{t:t+28} = 0$), while simultaneously accounting for reporting noise and imported cases.

We introduce infection incidence as an additional hidden state, which is assumed to evolve according to the renewal model. The expected number of local infections $I_t$ at time $t$ is a function of $R_t$, the PMF of the generation time distribution $\omega_u$, past local infections $I_{1:t-1}$, and past imported cases $M_{1:t-1}$:

\begin{equation}
    \begin{aligned}
        \log R_t | \log R_{t-1} &\sim \text{Normal}(\log R_{t-1}, \sigma)\\
        I_t | R_t, I_{1:t-1} &\sim \text{Poisson}\left(R_t \sum_{u=1}^{u_{max}} (I_{t-u} + M_{t-u}) \omega_u\right)\\
    \end{aligned}
\end{equation}

\noindent For simplicity, we also represent the generation time PMF by $\{\omega_u\}_{u=1}^{u_{max}}$ and use the same distribution as the last example. In practice, this should be changed to reflect that we are now modelling infection-to-infection rather than case reporting-to-case reporting.

We then assume that reported cases $C_t$ follow a negative binomial distribution with mean $I_t$ and variance $I_t + \phi I_t^2$:

\begin{equation}
    C_t | I_{1:t-1} \sim \text{Negative Binomial}\left(r = \frac{1}{\phi}, p = \frac{1}{1 + \phi I_t}\right)
\end{equation}

\noindent Parameter $\phi$ controls the level of observation noise, which we estimate alongside $\sigma$ using PMMH. Like $\sigma$, we use a uniform prior distribution on $(0, 1)$ for $\phi$.

The probability of elimination at time $t$ is calculated by projecting the hidden states forward in time, conditioning on $M_{t:t+28} = 0$ (no imported cases), and calculating the proportion of particle trajectories that contain zero new local infections. This is repeated at each time step to produce a time series of the probability of elimination.

Using PMMH (Algorithm \ref{alg:pmmh}) we estimate a posterior mean and 95\% credible interval of $\sigma$ of 0.16 (95\% Cr.I. 0.09, 0.25) and $\phi$ of 0.018 (95\% Cr.I. 0.0005, 0.065). The decreased estimate of $\sigma$ (compared to model 1) reflects the re-attribution of some noise in the data from the epidemic process to the observation process. The estimate of $\phi$ is small, suggesting that the observation process may be adequately modelled by a Poisson distribution, instead of the negative binomial distribution we have assumed (this could be tested using the model selection metrics outlined in Section \ref{sec:modelselection}). Convergence was obtained in 800 iterations, taking approximately 50 seconds on a 2021 M1 MacBook Pro.

\begin{figure}[h!]
    \centering
    \includegraphics[width=\textwidth]{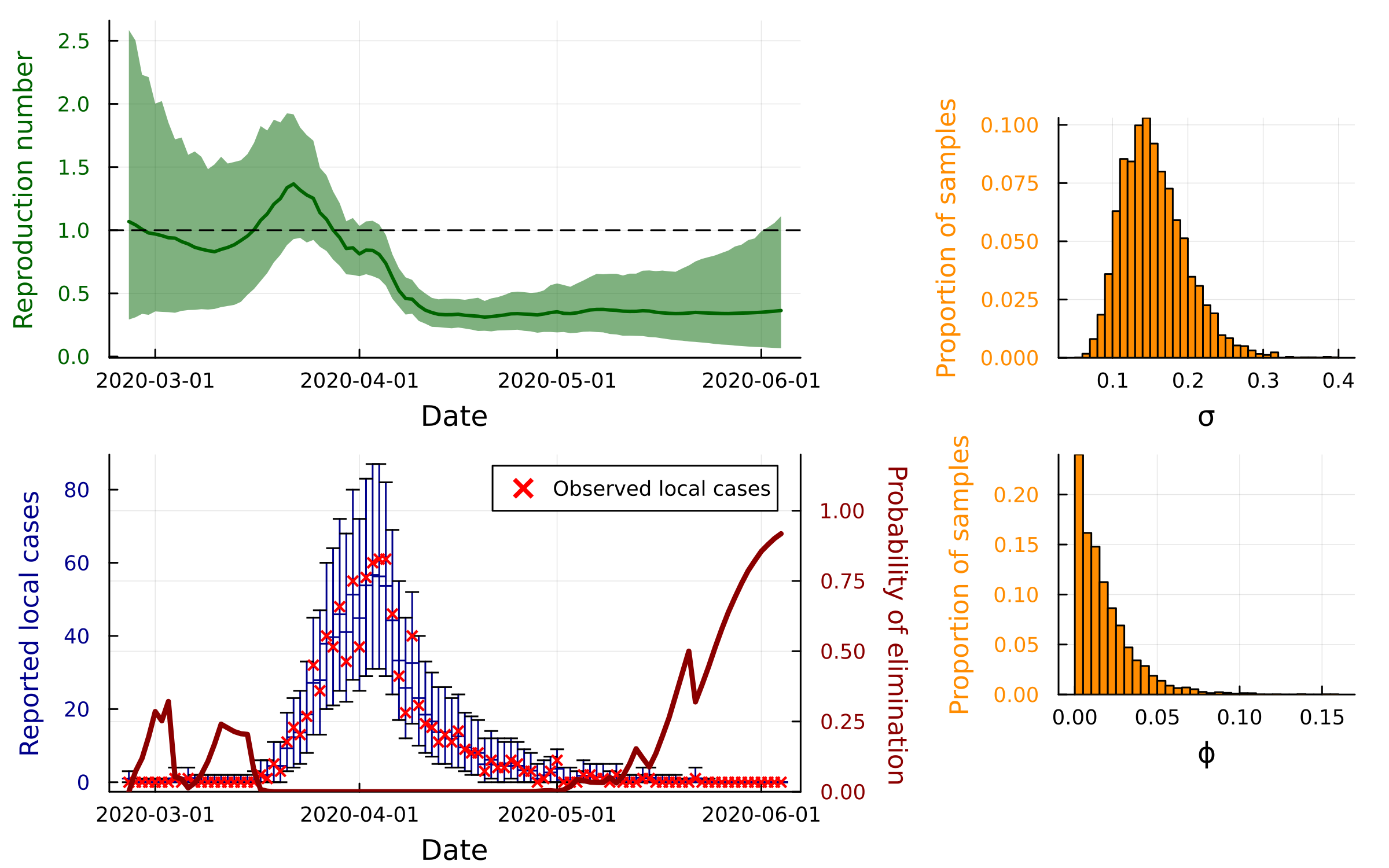}
    \caption{Results from fitting model 2 to data from the COVID-19 pandemic in New Zealand between 26 February 2020 and 4 June 2020. The marginal posterior means and 95\% credible intervals of $R_t$ (panel A) demonstrate a much lower and somewhat later peak in March 2020 than implied by model 1, reflecting the re-attribution of noise to the observation process and the distinction between locally-acquired and imported cases. The marginal posterior predictive distributions of reported cases (panel B) show that the model fits the observed data well, with observed data often lying close to posterior predictive means, although the model is overcovering the observed data. The probability of elimination increases steadily from early May and is estimated to reach 84.3\% on the final time step 4 June 2020. The histogram of parameter samples (panels C and D) show a high posterior probability that $\phi$ is near 0, suggesting that the observation process may be adequately modelled by a Poisson distribution instead of the overdispersed negative binomial distribution.}
    \label{fig:noiseimportselim}
\end{figure}

Daily posterior means and 95\% credible intervals of $R_t$ are reported in Figure \ref{fig:noiseimportselim}-A. Despite being fit to data from the same outbreak as model 1, the estimates of $R_t$ are very different. By halting the algorithm on 5 April 2020, we estimate that $R_t$ peaked at 1.6 (95\% Cr.I. 1.2, 2.2) on 23 March 2020 (95\% Cr.I. 6 March, 27 March), much lower and somewhat later than estimates from the simple model. This decrease is primarily a result of distinguishing locally-acquired from imported cases, although the re-attribution of noise from the epidemic process to the observation process also plays a role.

Figure \ref{fig:noiseimportselim}-B demonstrates that the model fits the data well, although 100\% of observed cases fall inside the 95\% posterior predictive credible intervals, suggesting that the model is allowing for too much observation noise. Plotting the histogram of posterior samples of $\phi$ (Figure \ref{fig:noiseimportselim}-D, from an extended implementation of PMMH with a minimum ESS of 1000) suggests that $\phi = 0$ (i.e. Poisson observation noise) is a plausible value. The uniform prior distribution places little mass around $\phi = 0$, which may be the cause of the overly-wide posterior predictive credible intervals.

The probability of elimination increases steadily from early May and is estimated to reach 91.8\% on the final time step 4 June 2020. Conditioning on no new imported cases is equivalent to assuming that any imported infections cause no local transmission, a reasonable assumption given the strict border controls in place at the time.

\subsection{Model 3: Reporting biases, temporal effects, and projections}
\label{sec:temporalmodels}

For the final example, we consider reported cases of COVID-19 in New Zealand between 1 April 2024 and 9 July 2024, a period characterised by widespread transmission, decreased testing rates, and increased reporting delays and noise. A pronounced day-of-week effect is evident in the data, with fewer cases reported on weekends. This can be accounted for by explicitly including a day-of-week effect or by modelling temporally aggregated data. We demonstrate both here.

We use a similar state-space transition model as the previous example, although ignore imported cases for simplicity:

\begin{equation}
    \begin{aligned}
        \log R_t | \log R_{t-1} &\sim \text{Normal}(\log R_{t-1}, \sigma)\\
        I_t | R_t, I_{1:t-1} &\sim \text{Poisson}\left(R_t \sum_{u=1}^{u_{max}} I_{t-u} \omega_u\right)\\
    \end{aligned}
\end{equation}

To account for reporting delays, the expected number of cases $\mu_t$ at time $t$ is modelled as a function of past incidence $I_{1:t-1}$ and an incubation period PMF $d_u$, assumed to be Gamma-distributed with mean 5.5 days and standard deviation 2.3 days (an incubation period distribution previously used for COVID-19 modelling in New Zealand\cite{hendyMathematicalModellingInform2021}, also discretised by evaluating at $t = 1, 2, \ldots$):

\begin{equation}
    \mu_t = \sum_{u=1}^{u_{max}} I_{t-u} d_u
\end{equation}

\noindent The ability to handle a wide range of approaches to modelling reporting delays is a key strength of the SMC approach. We discuss this further \href{https://nicsteyn2.github.io/SMCforRt/models_obsnoise.html}{online}.

For the day-of-week model, we introduce seven new parameters: $c_i, i = 1, \ldots, 7$, representing the relative reporting rate on day $i$. These parameters are subject to the constraint that $\sum_{i=1}^7 c_i = 7$, enforced by estimating $c_1, \ldots, c_6$ and setting $c_7 = 7 - \sum_{i=1}^6 c_i$. If $c_i > 1$, then more cases are reported on day $i$ than on average (and vice-versa). The observation distribution in this case is given by:

\begin{equation}
    C_t | \mu_t, \phi \sim \text{Negative Binomial}\left(r = \frac{1}{\phi}, p = \frac{1}{1 + \phi c_{\text{mod}(t, 7)+1} \mu_t}\right)
\end{equation}

\noindent where $c_{\text{mod}(t, 7)+1}$ is the day-of-week effect for day $t$. The $\text{mod}(t,7)$ term is the modulo operator, which allows us to cycle through the days of the week.

For the temporally-aggregated model, we calculate particle weights and resample on a weekly basis. When $\text{mod}(t, 7) = 0$, we define $C_t' = \sum_{i=t-6}^t C_i$ as the non-overlapping weekly aggregated data, and the observation distribution is given by:

\begin{equation}
    C_t' | \mu_{t-6:t}, \phi \sim \text{Negative Binomial}\left(r = \frac{1}{\phi}, p = \frac{1}{1 + \phi \sum_{i=t-6}^t \mu_i}\right), \ \ \ \text{if } \text{mod}(t, 7) = 0
\end{equation}

We compare these models to a third ``naive'' model, which fits to daily cases while ignoring the day-of-week effect, using the following observation distribution:

\begin{equation}
    C_t | \mu_t, \phi \sim \text{Negative Binomial}\left(r = \frac{1}{\phi}, p = \frac{1}{1 + \phi \mu_t}\right)
\end{equation}

\begin{figure}[h!]
    \centering
    \includegraphics[width=\textwidth]{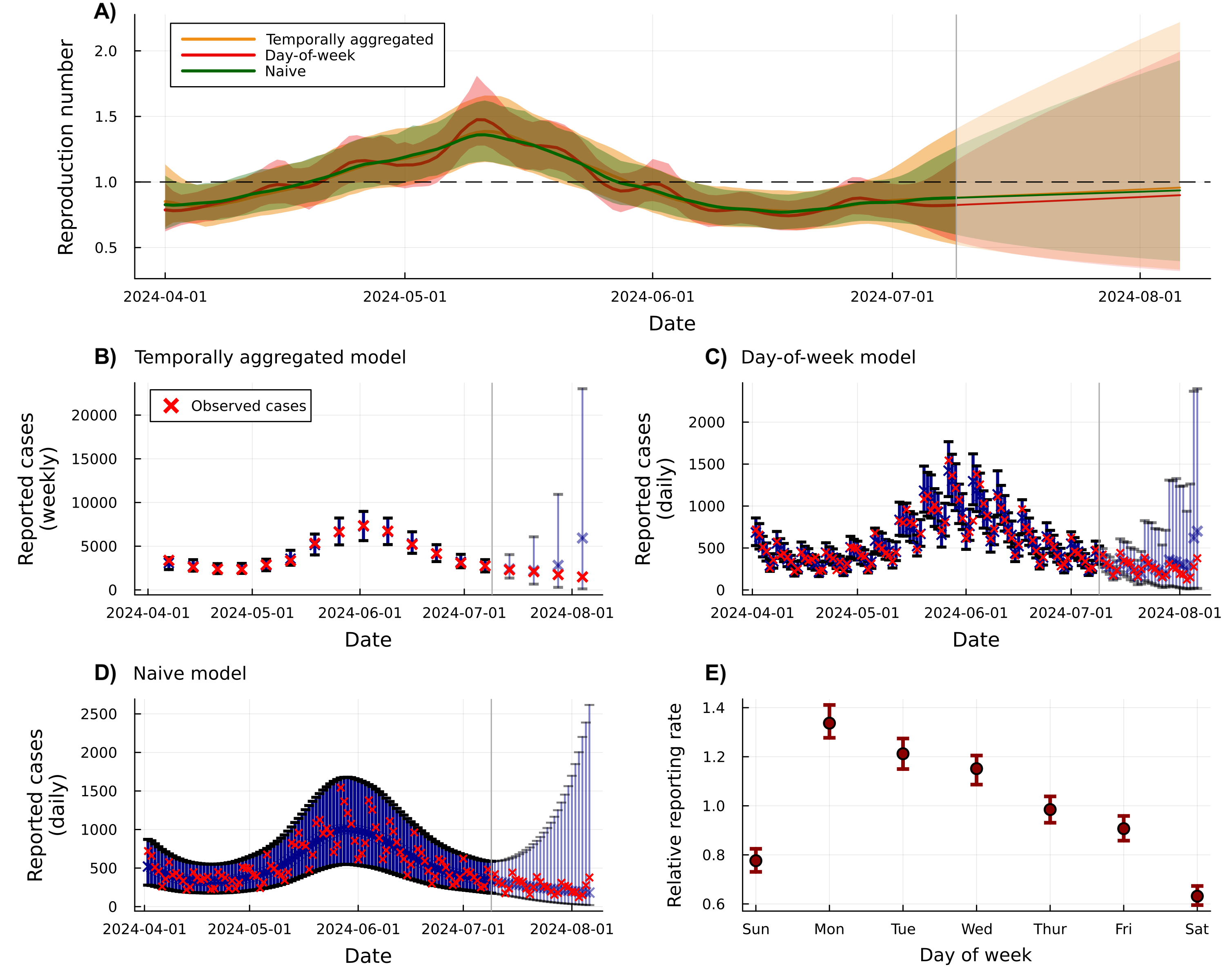}
    \caption{Results from fitting the three models of model 3 to data from the COVID-19 pandemic in New Zealand between 1 April 2024 and 9 July 2024. The dark red line and shaded region present the marginal posterior means and 95\% credible intervals of $R_t$ for the day-of-week model (red), the temporally aggregated model (orange), and naive model in green (panel A). Predictive weekly cases from the temporally aggregated model are shown in panel B, predictive daily cases from the day-of-week model are shown in panel C, and predictive daily cases from the naive model are shown in panel D. The day-of-week relative reporting rate estimates and 95\% credible intervals are shown in panel E. Projections of $R_t$ and reported cases are shown in panels (A-D), with lighter shading and the vertical grey bar marking the start of the projection.}
    \label{fig:temporalmodels}
\end{figure}

Using PMMH (Algorithm \ref{alg:pmmh}), we find that all three models produce similar estimates of $\sigma$ (Table \ref{tab:temporalmodels}). The day-of-week and temporally aggregated models produce comparable estimates of $\phi$, while the estimated overdispersion in the naive model is much higher. This is expected, as the naive model attributes day-of-week noise to overdispersion in the observation process, while the other models explicitly account for this. Estimates of the day-of-week relative reporting rates from the day-of-week model are given in Figure \ref{fig:temporalmodels}-E, highlighting statistically significantly greater reporting rates on weekdays (particularly Monday) than on weekends.

\begin{table}[h!]
    \centering
    \caption{Parameter estimates from fitting the day-of-week model, the temporally aggregated model, and the naive model to data from the COVID-19 pandemic in New Zealand between 1 April 2024 and 9 July 2024. 95\% credible intervals are shown in parenthesis. Estimates of the day-of-week relative reporting rates are shown in Figure \ref{fig:temporalmodels}-E.}
    \label{tab:temporalmodels}
    \begin{tabular}{l|cc}
        \toprule
        Model & $\sigma$ & $\phi$\\
        \midrule
        Day-of-week & 0.074 (0.043, 0.132) & 0.011 (0.007, 0.015)\\
        Temporally aggregated & 0.070 (0.039, 0.132) & 0.008 (0.001, 0.038)\\
        Naive & 0.065 (0.039, 0.116) & 0.073 (0.053, 0.092)\\
        \bottomrule
    \end{tabular}
\end{table}

Figure \ref{fig:temporalmodels}-A presents the daily posterior mean and 95\% credible intervals of $R_t$ for all three models. All three models produce similar estimates of $R_t$, although the day-of-week model exhibits slightly less uncertainty in some periods, reflecting the increase in information extracted from the data. This model is also able to capture the day-of-week effect while maintaining good statistical coverage, with 96.9\% of observed cases falling inside the 95\% posterior predictive credible intervals. The temporally aggregated and naive models both overcover the observed data, with 100\% of observations falling inside the 95\% predictive credible intervals, although there are only 14 observations in the temporally aggregated example.

Four-week projections are produced by iterating the estimated state-space model forward and sampling from the observation model. In all three models, 100\% of future cases fall inside the 95\% posterior predictive credible intervals, suggesting that all models overestimate the level of uncertainty (producing credible intervals that are too wide), most likely a result of model misspecification. Allowing for more complex correlation structures in the dynamic model for $R_t$ may reduce this uncertainty (for example, the assumption that $R_t$ follows a random walk ignores potential mean-reverting behaviour) \cite{banholzerComparisonShorttermProbabilistic2023}.

As we fit the naive model and the day-of-week model to the same data, we can compare the two using RMSE and CRPS. The RMSE of the naive model is 5.4 times greater than the day-of-week model on within-sample predictions and 255 times greater on the 4-week projections, suggesting that the day-of-week model produces much more accurate central estimates. CRPS, which simultaneously accounts for calibration and precision, is 2.6 times greater for the naive model for within-sample predictions and 1.53 times greater on out-of-sample (four-week ahead) predictions. Despite its additional complexity and similar hidden-state estimates, the day-of-week model substantially outperforms the naive model.

\section{Discussion}

We have introduced a general framework for fitting discrete-time epidemic renewal models using SMC methods and demonstrated their application in estimating hidden states, parameters, and making projections. The primary strength of these methods lies in their flexibility: they can fit any state-space model, a form that encompasses many popular epidemiological models. This allows for rapid fitting, testing, and comparison of a wide range of models, facilitating the rapid development of solutions to emerging epidemiological challenges. All code and data to reproduce all results in this paper (and more) are provided \href{https://nicsteyn2.github.io/SMCforRt/}{online}.

We demonstrate our methods on three examples. Each example consists of a distinct model structure, dataset, and aim, yet the same general framework is used to fit each model. The first example demonstrates the estimation of $R_t$ using a simple renewal model, similar to EpiFilter \cite{paragImprovedEstimationTimevarying2021}, with the addition of joint estimation of the peak $R_t$ value and the timing of this peak. The second example demonstrates the estimation of $R_t$ while accounting for both reporting noise \cite{whiteReportingErrorsInfectious2010, sherrattExploringSurveillanceData2021} and imported cases \cite{thompsonImprovedInferenceTimevarying2019, robertsEarlyEstimationReproduction2011, churcherMeasuringPathMalaria2014}, while simultaneously estimating the probability of elimination \cite{paragDecipheringEarlywarningSignals2021, plankEstimationEndofoutbreakProbabilities2025}, thus unifying the cited works. The final example demonstrates the estimation of $R_t$ in the presence of reporting delays \cite{lawlessAdjustmentsReportingDelays1994, azmonEstimationReproductionNumber2014, sherrattExploringSurveillanceData2021} and day-of-week effects \cite{azmonEstimationReproductionNumber2014, alvarezRemovingWeeklyAdministrative2020} or temporally aggregated data \cite{ogi-gittinsEfficientSimulationbasedInference2024, ogi-gittinsSimulationbasedApproachEstimating2024, nashEstimatingEpidemicReproduction2023}, as well as the ability to produce short-term projections \cite{nouvelletSimpleApproachMeasure2018}, and the importance of model comparison and selection \cite{banholzerComparisonShorttermProbabilistic2023, gosticPracticalConsiderationsMeasuring2020}, again unifying the cited (and many non-cited) works. Many other extensions are possible, particularly the ability to incorporate multiple data sources \cite{deangelisFourKeyChallenges2015, watsonJointlyEstimatingEpidemiological2024}.

Many modelling choices made in our examples are arbitrary: for example, a Poisson observation distribution could be used in the second example (instead of a negative binomial distribution), the day-of-week-effect could be modelled as daily binomial reporting probabilities, and there are many ways to define elimination, to name just a few. Different researchers address the same problems in different ways. The core strength of these methods is their ability to quickly and easily test these different models. We also provide principled methods for model evaluation and selection, including RMSE, CRPS, and coverage of posterior predictive credible intervals. These metrics are useful for comparing models fit to the same data, but less so when models are fit to different data sources, a potential area for future work.

Correctly accounting for uncertainty in model parameters has previously been shown to be critical for robust uncertainty quantification, and thus for robust policymaking \cite{steynRobustUncertaintyQuantification2024, gressaniEpiLPSFastFlexible2022}. Our focus on marginalising out parameter uncertainty is a key strength of these methods, and is not included in many other epidemiological methods, including many SMC-based approaches \cite{yangBayesianDataAssimilation2022, plankEstimationEndofoutbreakProbabilities2025}. By coupling existing particle filter-only approaches with a PMMH algorithm like that presented here, an additional source of uncertainty (uncertainty in fixed parameters) can be accounted for, increasing the robustness of these methods.

Epidemic models range from highly mechanistic (such as compartmental SEIR-type models) to purely statistical (such as time series regression or exponential smoothers). The renewal model is semi-mechanistic in that it imposes a simple structure but does not model the entire underlying process. This leads to a model that is flexible, interpretable, and can produce well-calibrated short-term projections. However, like any mechanistic model, the renewal model and dynamic model on $R_t$ jointly imply a specific autocorrelation structure in the data. If this structure is misspecified and the only goal is the projection of observable data, then model may underperform compared to more flexible statistical models \cite{banholzerComparisonShorttermProbabilistic2023}.

Our focus on simplicity and flexibility results in algorithms that are known to be less computationally efficient than more specialised alternatives. Areas for additional consideration include the use of a better-tuned proposal distribution (in Algorithm 1) \cite{pittFilteringSimulationAuxiliary1999}, the use of more advanced resampling schemes (in Algorithm 1) \cite{chopinIntroductionSequentialMonte2020}, and the use of better-tuned PMMH proposal densities (in Algorithm 2) \cite{kantasParticleMethodsParameter2015, hurzelerApproximatingMaximisingLikelihood2001}. Many solutions to these problems exist within the literature, often leveraging specific details about the chosen model, and we encourage the reader to seek out more efficient algorithms once a suitable model has been specified. In addition to epidemiology-specific literature \cite{temfackReviewSequentialMonte2024b,endoIntroductionParticleMarkovchain2019}, we also recommend two texts from the broader SMC literature \cite{doucetSequentialMonteCarlo2001, chopinIntroductionSequentialMonte2020} and a tutorial paper by Doucet \& Johansen (2011)\cite{doucetTutorialParticleFiltering2011}.

An effective class of alternatives to SMC-type methods for performing inference and generating projections with epidemic renewal models are probabilistic-programming-language-based Gaussian process approximations, such as EpiNow2 \cite{abbottEstimatingTimevaryingReproduction2020}. These methods offer a similar level of flexibility and, in some cases, can be faster than our approach. However, these methods are more complex mathematically, whereas ours are more accessible to those without a strong mathematical background and can be implemented in a few lines of code, making them easier to debug. Furthermore, our methods can easily be adapted for online updates, allowing for real-time analysis of epidemic data, whereas Gaussian-process-based models are typically offline methods. Finally, popular probabilistic programming languages, such as Stan, do not support discrete parameters. Many quantities of interest, such as infection incidence, are discrete, which our methods can handle natively, whereas other approaches require additional approximations.

Numerous methods for estimating $R_t$ and conducting inference on epidemic data exist, often tailored to specific pathogens or datasets. More general models have also been created, although these can be challenging to implement or are confined to pre-built software packages. While these existing packages simplify the process, they can obscure the underlying structure of the model and associated assumptions. This primer aims to strike a balance: enabling researchers to quickly and easily construct their own models from scratch, ensuring they understand the assumptions and structure of the model, while facilitating rapid testing and comparison of different models. By providing a general framework for fitting discrete-time epidemic renewal models, we hope these methods contribute to the development of robust and useful epidemic models that can inform public health decision-making in future outbreaks across a range of pathogens.



\bmsection*{Acknowledgments}

We would like to thank Cathal Mills for helpful discussion and feedback throughout the development of this paper and the corresponding website, Ben Cooper and Ben Lambert for comments on an earlier version of this paper, and members of the Infectious Disease Modelling group at the Mathematical Institute in Oxford for helpful discussions throughout the project. We would also like to thank Alice Chen and Otto Arends Page for highlighting additional benefits of this framework when compared to existing methods.

\bmsection*{Financial disclosure}

N.S. acknowledges support from the Oxford-Radcliffe Scholarship from University College, Oxford, the EPSRC CDT in Modern Statistics and Statistical Machine Learning (Imperial College London and University of Oxford), and A. Maslov for studentship support. K.V.P. acknowledges funding from the MRC Centre for Global Infectious Disease Analysis (Reference No. MR/X020258/1) funded by the UK Medical Research Council. This UK-funded grant is carried out in the frame of the Global Health EDCTP3 Joint Undertaking. C.A.D. acknowledges support from the MRC Centre for Global Infectious Disease Analysis, the NIHR Health Protection Research Unit in Emerging and Zoonotic Infections, the NIHR funded Vaccine Efficacy Evaluation for Priority Emerging Diseases (PR-OD-1017-20007), and the Oxford Martin Programme in Digital Pandemic Preparedness. The funders had no role in study design, data collection and analysis, decision to publish, or manuscript preparation.

\bmsection*{Supporting information}

All code required to reproduce these results, and further discussion on a range of topics, is provided online at \href{https://nicsteyn2.github.io/SMCforRt/}{https://nicsteyn2.github.io/SMCforRt/}.



\nolinenumbers

\bibliography{bib}

\end{document}